\documentclass[10pt]{article}
\usepackage[latin1]{inputenc}
\usepackage{amsmath}
\usepackage{amsfonts}
\usepackage{amssymb}
\usepackage{mathrsfs}

\usepackage[cal=boondox,scr=boondoxo]{mathalfa}

\usepackage{float}
\usepackage{subfig}
\usepackage{fancybox,graphicx}
\usepackage{subfig}
\usepackage{caption}
\usepackage{color}
\usepackage{authblk}

\usepackage[colorlinks]{hyperref}
\usepackage{hyperref}

\newcommand{\doi}[1]{{doi:~\href{https://doi.org/#1}{\nolinkurl{#1}}}\rmFullStop}

\newcommand*{\rmFullStop}{\rmifnextchar{.}{}{}}

\makeatletter
\newcommand{\rmifnextchar}[3]{%
  \begingroup
  \ltx@LocToksA{\endgroup#2}%
  \ltx@LocToksB{\endgroup#3}%
  \ltx@ifnextchar{#1}{%
    \def\next{\the\ltx@LocToksA}%
    \afterassignment\next
    \let\scratch= %
  }{%
    \the\ltx@LocToksB
  }%
}
\makeatother %doi command

\usepackage{accents}
\usepackage[titletoc,title]{appendix}
\usepackage{cite}

\usepackage{mathtools}

\DeclarePairedDelimiter\floor{\lfloor}{\rfloor}

\usepackage[top=2in, bottom=1.5in, left=1in, right=1in]{geometry}

\title{Evaluation of the Biot-Savart integral in electrostatic problems with non-uniform Dirichlet boundary conditions} 
 
\author[1,3]{Robert Salazar}
\author[1,2]{Camilo Bayona}
\author[1]{J. S. Sol\'is Chaves}
\affil[1]{Universidad ECCI - Bogot\'a, Colombia}
\affil[2]{Centro de Ingenier\'ia Avanzada Investigaci\'on y Desarrollo, CIAID- Bogot\'a, Colombia}
\affil[3]{Departamento de F\'isica, Universidad de los Andes - Bogot\'a, Colombia}
\begin{document}

    \maketitle

\begin{abstract}
We present an analytical strategy to solve the electric field generated by a planar region which is kept with a fixed but non-uniform electric potential. The approach can be used in certain situations where the electric potential on the space requires to solve the Laplace equation with non-uniform Dirichlet boundary conditions. We show that the electric field is due to a contribution depending on the circulation on the contour in a Biot-savart way plus another one taking into account the angular variations of the potential in $\mathcal{A}$ valid for any closed loop $c$. The approach is used to find exact expansions solutions of the electric field for circular contours with fully periodic potentials. Analytical results are validated with numerical computations and the Finite Element Method.\\\\Keywords: Biot-Savart law, electrostatic problems, exactly solvable models.  
\end{abstract}

\section{Introduction}
The computation of the electric field generated by a stationary charge density distribution is a standard problem of physics and engineering. Conceptually, the problem is simple, however its exact solution can be difficult depending on how the charge density is distributed in space. In principle, it is possible to find analytic solutions to the problem when the charge density is uniform \cite{hummer1996electrostatic,lekner2010analytical,ciftja2013calculation,polyakov2015new,mccreery2018electric}. A particular situation occurs when a fixed but non uniform potential is distributed on a boundary surface of the domain. These type of Dirichlet problems implies a charge density on the surface, and the electric potential due to this density charge can be obtained by solving a Laplace's equation. At best, the systems is integrable, as it occurs with some axially symmetric problems on the sphere \cite{jackson1999classical,griffiths2005introduction} and the disk \cite{atkinson1983analytic}.
In those cases, standard strategies of separation of variables can be used to find the electric potential. However, the computation of the electric potential for more complicated geometries often requires numerical approximations as the only possible approach.        

\begin{figure}[h]
\centering %system.pdf
\includegraphics[width=0.6\textwidth]{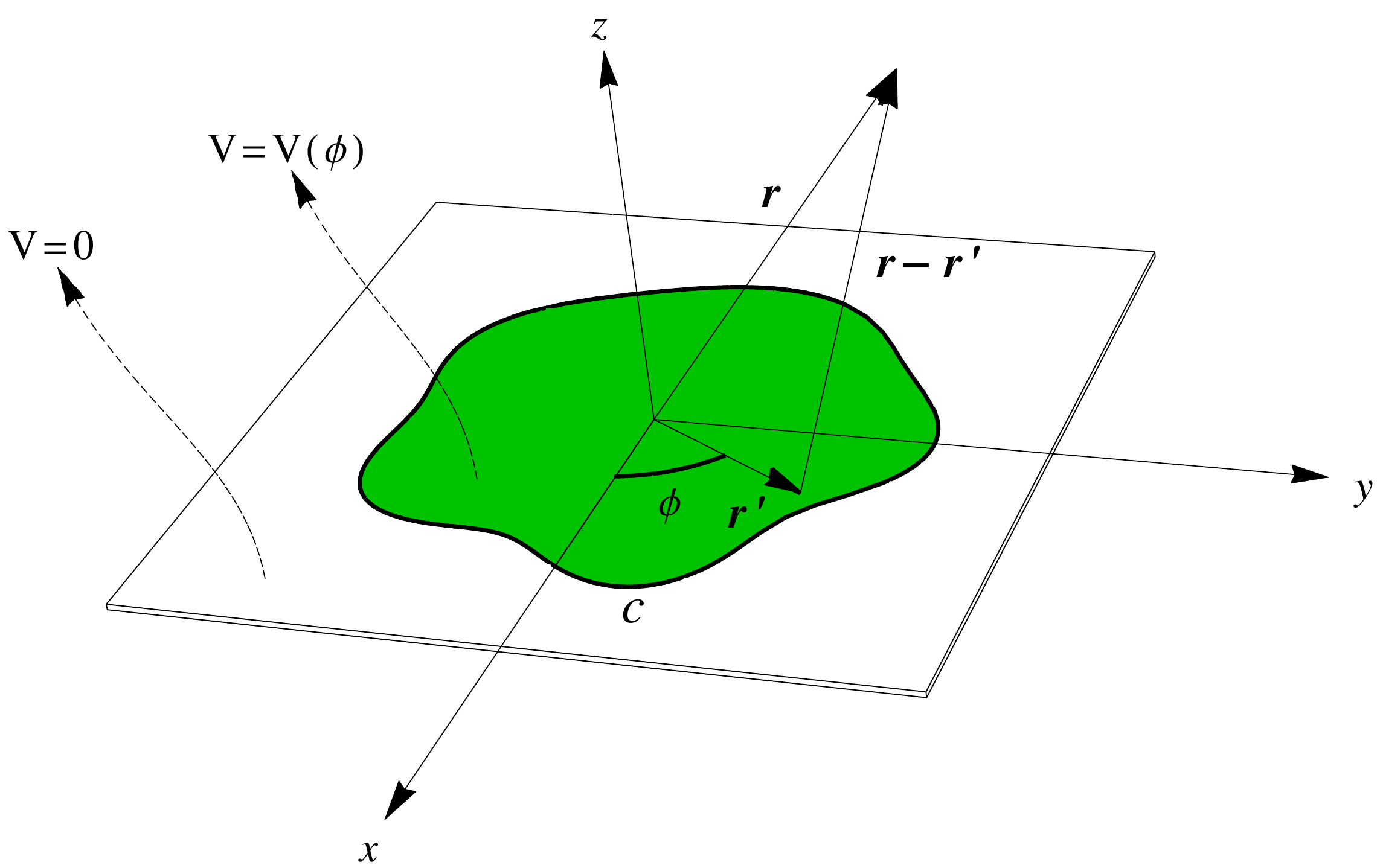}
    \caption[The system.]{Planar region of arbitrary contour $c$ with a $\phi$-dependent electric potential $V(\phi)$}
\label{theSystemFig}
\end{figure}

In this document we shall describe a technique to compute the electric field of planar region with a non-uniform electric potential with the rest of the plane grounded, as it is shown in Fig.~\ref{theSystemFig}. Using a cylindrical reference system, we define $\boldsymbol{r}=(r,\phi,z)$ to be any position inside the domain $\mathfrak{D}=\left\{\boldsymbol{r} \in \mathbb{R}^3 : z \geq 0 \right\}$. A non-uniform electric potential $V=V(\phi)$ is defined inside the planar region $\mathcal{A}$ which is located at $z=0$ and enclosed by the curve $c$. Although, this type of classical problems can be solved by using numerical methods \cite{shortley1938numerical,rangogni1986numerical,gray1986program, li2011finite}, the determination of analytic solutions for the electric field of this system is still a challenging problem. Nevertheless, there are significant simplifications when $V$ is held as a non-zero constant. In that case, the system exhibits a connection with magnetostatics and the Biot-Savart law \cite{oliveira2001biot}. The main objective of this work is to determine if this connection between electrostatics and magnetostatics still exist for arbitrary scalar potentials $V(\phi)$, and to use it to simplify the analytic problem.

This document will be organized as follows: In Section 2 we shall derive an expression to compute the electric field $\boldsymbol{E}(\boldsymbol{r})$ by using a connection between this electrostatic problem and magnetostatics. In that section we shall demonstrate that this expression is
\begin{equation}
\boldsymbol{E}(\boldsymbol{r})  = \frac{1}{2\pi} \oint_{ c } V(\phi') \frac{(\boldsymbol{r}-\boldsymbol{r}') \times d\boldsymbol{r}'}{|\boldsymbol{r}-\boldsymbol{r}'|^3} + \frac{1}{2\pi} \int_{0}^{2\pi}  V(\beta) \partial_\phi \boldsymbol{f}(\beta,\boldsymbol{r}) d\beta \hspace{0.5cm}\mbox{for}\hspace{0.5cm}z>0,
\label{centralResultEq}    
\end{equation}
where the line integral at the right side of the previous expression is analogous to the problem of computing the magnetic field along a loop $c$, but in a counter-intuitive situation where the loop carries a non-uniform electric current. The second term in the right of Eq.~$(\ref{centralResultEq})$ takes into account the electric field contributions due to variations of $V(\phi)$ in the region enclosed by $c$. The exact evaluation of Eq.~$(\ref{centralResultEq})$ for arbitrary contours can be difficult to evaluate, except for circular loops. Hence, in Section 3 we demonstrate the application of exact analytic expansions to solve Eq.~$(\ref{centralResultEq})$ in the case of circular contours with any arbitrary but periodic potential functions. The next section is devoted to the comparison of the analytic results with the numerical ones obtained by brute-force numerical integration. We also present in Section 4 some comparisons with the numerical solution given by the Finite Element Method. Finally, some conclusions are stated in Section 5.

\section{Electric field via Biot-Savart}
In general, a steady electric field $\boldsymbol{E}$ has associated an electric potential $\Phi(\boldsymbol{r})$ such that $\boldsymbol{E} = - \mbox{grad}\Phi(\boldsymbol{r})$. This, in combination with the Gauss's law, results in the Poisson's equation
\[
\nabla^2 \Phi(\boldsymbol{r}) = -\frac{\rho(\boldsymbol{r})}{\epsilon_o}\hspace{0.5cm}\mbox{for}\hspace{0.5cm} \boldsymbol{r} \in \mathfrak{D},
\]
where $\rho(\boldsymbol{r})$ is the volume charge density \cite{jackson1999classical,sadiku2014elements,griffiths2005introduction}.  We suppose that $\rho(\boldsymbol{r}) = 0$ for $z>0$, such that the Laplace equation 
\[
\nabla^2 \Phi(\boldsymbol{r}) = 0, \hspace{0.5cm} \boldsymbol{r} \in \mathfrak{D},
\]
subjected to the following boundary conditions
\[
\Phi(\boldsymbol{r}) = V(\phi)  \hspace{0.5cm} \mbox{\textbf{if}} \hspace{0.5cm} \boldsymbol{r} \in \mathcal{A} \subset \left\{\boldsymbol{r} \in \mathbb{R}^2 : z = 0 \right\}, \hspace{0.5cm}
\Phi(\boldsymbol{r}) = 0 \hspace{0.5cm} \mbox{\textbf{if}} \hspace{0.5cm} \boldsymbol{r} \in \left\{\boldsymbol{r} \in \mathbb{R}^2 : z = 0 \right\}\setminus\mathcal{A},
\]
has to be solved in the domain.
The electric potential can be obtained from the solution of the Poisson's equation using Green's functions $G(\boldsymbol{r},\boldsymbol{r}')$ \cite{jackson1999classical}. The solution is written as
\[
\Phi(\boldsymbol{r}) = \frac{1}{4\pi\epsilon_o}\int_{\mathfrak{D}} \rho(\boldsymbol{r}) G(\boldsymbol{r},\boldsymbol{r}')d^3 \boldsymbol{r}' + \frac{1}{4\pi}\oint_S \left[G(\boldsymbol{r},\boldsymbol{r}')\frac{\partial\Phi}{\partial n'} - \Phi(\boldsymbol{r'}) \frac{\partial G(\boldsymbol{r},\boldsymbol{r}')}{\partial n'} \right]d^2 \boldsymbol{r}',
\]
where the first term at the right hand vanishes since $\rho(\boldsymbol{r})=0$ for $r \in \mathfrak{D}$. As usual, we may demand that
\[
\left.G(\boldsymbol{r},\boldsymbol{r}')\right|_{z'=0} = 0,
\]
by choosing the Green's function for the half-space $z>0$ and considering the potential as a punctual charge at $(x',y',z')$, with $z'>0$, plus the potential of an image charge placed in a symmetric position in the lower-half plane, at $(x',y',-z')$. This Green's function results in
\[
G(\boldsymbol{r},\boldsymbol{r}') = \sum_{\sigma\in\{+1,-1\}}\frac{\sigma}{\sqrt{(x-x')^2+(y-y')^2+(z + \sigma z')^2}},
\]
Hence, the problem is reduced to evaluate the following integral
\begin{equation}
\Phi(\boldsymbol{r}) = - \frac{1}{4\pi}\int_{\mathcal{A}} V(\phi') \frac{\partial G(\boldsymbol{r},\boldsymbol{r}')}{\partial n'} d^2 \boldsymbol{r}',
\label{electricPotentialIntegralEq}
\end{equation}
where $\hat{n}'=-\hat{z}$ is the outward normal of $\partial\mathfrak{D}$ on the $(z=0)$-plane, and
\[
\left.\frac{\partial G(\boldsymbol{r},\boldsymbol{r}')}{\partial n'}\right|_{z=0} = - \left.\frac{2(z-z')}{|\boldsymbol{r}-\boldsymbol{r}'|^3}\right|_{z'=0},
\]
such that the electric potential in the cylindrical coordinate system can be calculated from
\begin{equation}
\Phi(r,\theta,\phi) = \frac{r\cos\theta}{2\pi} \int_{0}^{2\pi} \int_0^{\mathscr{R}(\phi)} \frac{V(\phi')r'dr' d \phi'}{(r^2+r'^2-2rr'\sin\theta\cos(\phi-\phi'))^{3/2}}.
\label{electricPotentialDoubleIntegralEq}
\end{equation}
In general, the previous expression of the electric potential for a given variable potential $V(\phi)$ is not easy to solve exactly and a numerical integration is needed. However, if the function $V(\phi)$ would be constant, say $V_o$, then, from Eq.~(\ref{electricPotentialIntegralEq}), the electric potential $\Phi_{unif.}(\boldsymbol{r})$ can be written as
\[
\Phi_{unif.}(\boldsymbol{r}) = \frac{V_o}{2\pi}\int_{\mathcal{A}} \frac{(\boldsymbol{r}-\boldsymbol{r}')\cdot \hat{z}}{|\boldsymbol{r}-\boldsymbol{r}'|^3} d^2 \boldsymbol{r}' = \frac{V_o}{2\pi} \Omega(\boldsymbol{r}).
\]
Authors of Ref.\cite{oliveira2001biot} noted that the physiscal quantity $\Omega(\boldsymbol{r})$ is proportional to the \textit{scalar potential} \cite{eyges2012classical,vanderlinde2006classical} of a steady magnetic field $\boldsymbol{B}(\boldsymbol{r})$ generated by electric current $i_o$ along a closed arbitrary loop $c$. In that problem, 
\[
\boldsymbol{B}(\boldsymbol{r}) = - \frac{\mu_o i_o}{4\pi} \nabla \Omega(\boldsymbol{r}) = \frac{\mu_o i_o}{4\pi} \oint_c  \frac{(\boldsymbol{r}-\boldsymbol{r}') \times d\boldsymbol{r}'}{|\boldsymbol{r}-\boldsymbol{r}'|^3}.
\]

\begin{figure}[H]
\centering %discreteSheet.pdf
\includegraphics[width=0.5\textwidth]{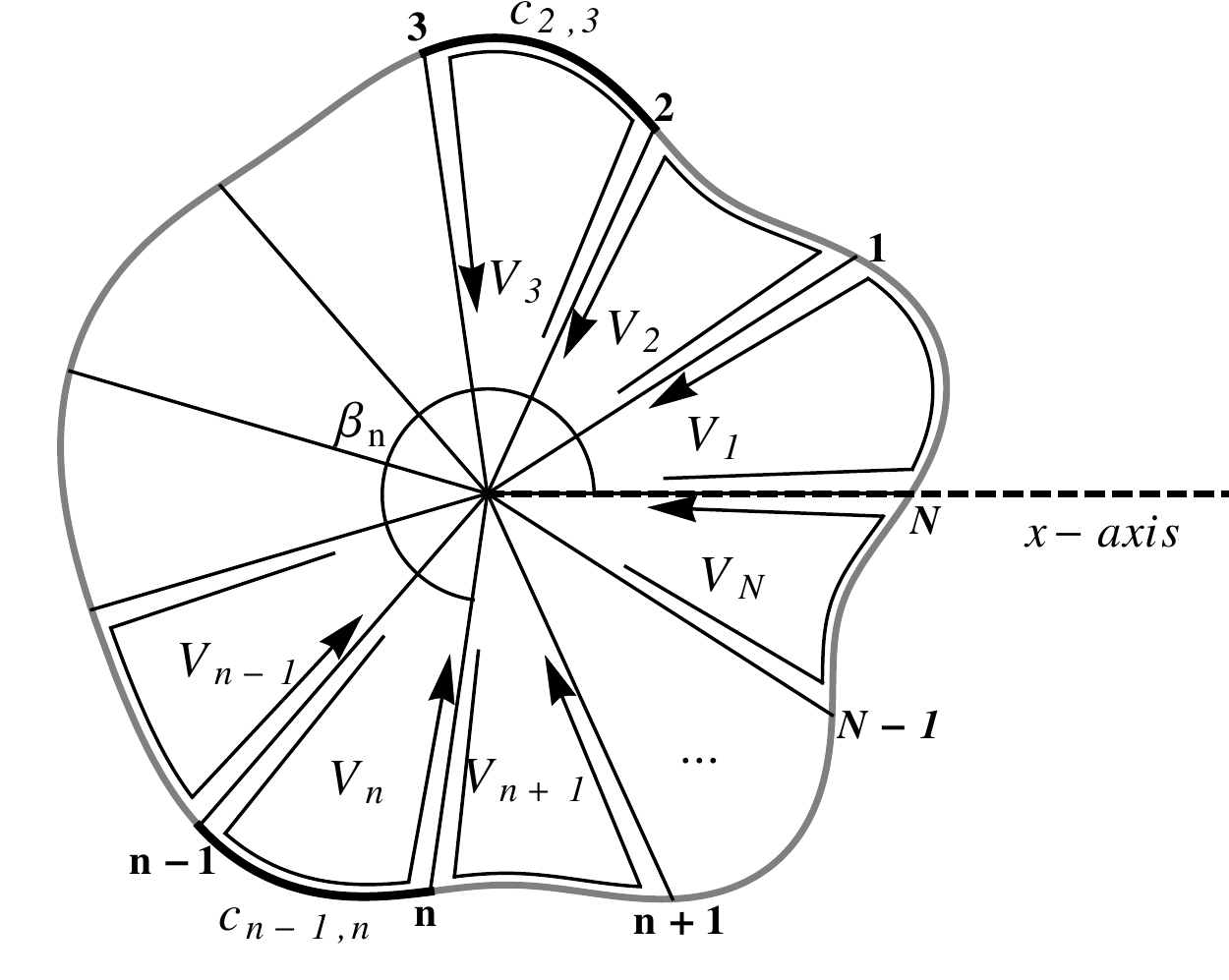}
    \caption{Discrete distribution of the potential on the sheet.}
\label{discreteSheetFig}
\end{figure}

Similarly,
\begin{equation}
\boldsymbol{E}_{unif.}(\boldsymbol{r}) = -\nabla \Phi_{unif.}(\boldsymbol{r})= - \frac{V_o}{2\pi} \nabla \Omega(\boldsymbol{r}) = \frac{V_o}{2\pi} \oint_c  \frac{(\boldsymbol{r}-\boldsymbol{r}') \times d\boldsymbol{r}'}{|\boldsymbol{r}-\boldsymbol{r}'|^3} \hspace{0.25cm}\mbox{for}\hspace{0.25cm}z>0,
\label{EUniformBiotEq}
\end{equation}
and therefore, the electric field can be computed by evaluating a Biot-Savart integral. Formally, Eq.~(\ref{EUniformBiotEq}) is only valid when $V$ is kept as constant and homogeneous in the region $\mathcal{A}$. The challenge is to generalize this result for any potential $V(\phi)$ with an arbitrary angular dependence. To this aim, let us consider the problem where $V(\phi)$ is not uniform but varies discretely as follows   
\[
V(\phi) = V_n  \hspace{0.5cm}\mbox{\textbf{if}}\hspace{0.5cm}\phi \in (\beta_{n-1},\beta_{n}),
\]
where $0<\beta_1 < \beta_2 < \ldots < \beta_N = 2\pi$ defines $N$ consecutive sectors $\{\mathcal{A}\}_{n=1,\dots,N}$ of uniform electric potential  $\left\{V_n\right\}_{n=1,\dots,N}$, where 
\[
\mathcal{A} = \mathcal{A}_1 \cup \mathcal{A}_2 \ldots \cup \mathcal{A}_N,
\]
as it is shown in Fig.~\ref{discreteSheetFig}. The boundary of each sector $A_n$ is the following loop
\[
c_n \cup c_{n-1,n} \cup \tilde{c}_{n-1}.
\]
Let us introduce an angular function $\mathscr{R}(\phi)$ that returns the curve's radius and helps to define the position of the external curve in polar coordinates
\[
c_{n-1,n} := \left\{ (\mathscr{R}(\phi), \phi) : \beta_{n-1} < \phi \leq \beta_n \right\}.
\]
Therefore, the complete closed curve $c$ can be defined as
\[
c = \left\{ (\mathscr{R}(\phi), \phi) : 0 < \phi \leq 2\pi \right\} = c_{1,2}\cup c_{2,3} \ldots \cup c_{N-1,N} \cup c_{N,1}.
\]
The path $c_n$ is a straight line from the point $(\mathscr{R}(\beta_n), \beta_n)$ to the origin and $\tilde{c}_n$ is its reversed trajectory. Since the potential of each sector is constant and homogeneous, we may use the superposition principle in Eq.~(\ref{EUniformBiotEq}) to compute the electric field in the domain as follows,
\[
\boldsymbol{E}(\boldsymbol{r}) = \sum_{n=1}^N \boldsymbol{E}_n(\boldsymbol{r})= \frac{1}{2\pi} \sum_{n=1}^N \int_{c_n \cup c_{n-1,n} \cup \tilde{c}_{n-1} } V_{n} \frac{(\boldsymbol{r}-\boldsymbol{r}') \times d\boldsymbol{r}'}{|\boldsymbol{r}-\boldsymbol{r}'|^3},
\]
where $\boldsymbol{E}_n(\boldsymbol{r})$ is the electric potential due to the $n$-th sector of $\mathcal{A}$. It is possible to split the integral term of the previous equation, resulting in
\[
\int_{c_n \cup c_{n-1,n} \cup \tilde{c}_{n-1} } V_{n} \frac{(\boldsymbol{r}-\boldsymbol{r}') \times d\boldsymbol{r}'}{|\boldsymbol{r}-\boldsymbol{r}'|^3} = \int_{c_{n-1,n}} V_{n} \frac{(\boldsymbol{r}-\boldsymbol{r}') \times d\boldsymbol{r}'}{|\boldsymbol{r}-\boldsymbol{r}'|^3} +  \int_{c_n} V_{n} \frac{(\boldsymbol{r}-\boldsymbol{r}') \times d\boldsymbol{r}'}{|\boldsymbol{r}-\boldsymbol{r}'|^3} - \int_{c_{n-1} } V_{n} \frac{(\boldsymbol{r}-\boldsymbol{r}') \times d\boldsymbol{r}'}{|\boldsymbol{r}-\boldsymbol{r}'|^3} .
\]
Thus, the electric field expression takes the form
\begin{equation}
\boldsymbol{E}(\boldsymbol{r}) = \frac{1}{2\pi} \int_{ \bigcup_{n=1}^N c_{n-1,n} } V(\phi') \frac{(\boldsymbol{r}-\boldsymbol{r}') \times d\boldsymbol{r}'}{|\boldsymbol{r}-\boldsymbol{r}'|^3} + \frac{1}{2\pi} \sum_{n=1}^N V_{n}  \left[ \int_{c_n} \frac{(\boldsymbol{r}-\boldsymbol{r}') \times d\boldsymbol{r}'}{|\boldsymbol{r}-\boldsymbol{r}'|^3} - \int_{c_{n-1} }  \frac{(\boldsymbol{r}-\boldsymbol{r}') \times d\boldsymbol{r}'}{|\boldsymbol{r}-\boldsymbol{r}'|^3} \right].
\label{auxForEFieldEq}
\end{equation}
The first term on the right of Eq.~(\ref{auxForEFieldEq}) is a circulation on the whole loop since $c=\bigcup_{n=1}^N c_{n-1,n}$. On the other hand, the sum of the straight lines' integrals can be written in a most convenient form by noting that
\[
\sum_{n=1}^N V_{n} \int_{c_{n-1} }  \frac{(\boldsymbol{r}-\boldsymbol{r}') \times d\boldsymbol{r}'}{|\boldsymbol{r}-\boldsymbol{r}'|^3} = - V_1 \int_{c_{0} } \frac{(\boldsymbol{r}-\boldsymbol{r}') \times d\boldsymbol{r}'}{|\boldsymbol{r}-\boldsymbol{r}'|^3} +  \sum_{n=1}^N V_{n+1} \int_{c_{n} } \frac{(\boldsymbol{r}-\boldsymbol{r}') \times d\boldsymbol{r}'}{|\boldsymbol{r}-\boldsymbol{r}'|^3} . 
\]
Therefore,
\[
\begin{split}
\boldsymbol{E}(\boldsymbol{r})  = & \frac{1}{2\pi} \oint_{ c } V(\phi') \frac{(\boldsymbol{r}-\boldsymbol{r}') \times d\boldsymbol{r}'}{|\boldsymbol{r}-\boldsymbol{r}'|^3} + \\ & \frac{1}{2\pi}  \left[ - V_1 \int_{c_{0} } \frac{(\boldsymbol{r}-\boldsymbol{r}') \times d\boldsymbol{r}'}{|\boldsymbol{r}-\boldsymbol{r}'|^3} + \sum_{n=1}^{N-1} (V_{n}-V_{n+1})\int_{c_n} \frac{(\boldsymbol{r}-\boldsymbol{r}') \times d\boldsymbol{r}'}{|\boldsymbol{r}-\boldsymbol{r}'|^3}  + V_N \int_{c_{N} } \frac{(\boldsymbol{r}-\boldsymbol{r}') \times d\boldsymbol{r}'}{|\boldsymbol{r}-\boldsymbol{r}'|^3}\right],
\end{split}
\]
and using the periodicity conditions
\begin{equation}
c_m = c_{N+m}, \hspace{0.5cm}\mbox{and}\hspace{0.5cm} V_{N+m} = V_{m},  \hspace{0.5cm}\forall\hspace{0.5cm}  m\in \mathbb{Z}^0,  
\label{periodicityConditionsEq}
\end{equation}
then the electric field can be written as
\begin{equation}
\boldsymbol{E}(\boldsymbol{r})  = \frac{1}{2\pi} \oint_{ c } V(\phi') \frac{(\boldsymbol{r}-\boldsymbol{r}') \times d\boldsymbol{r}'}{|\boldsymbol{r}-\boldsymbol{r}'|^3} + \frac{1}{2\pi}  \sum_{n=1}^N (V_{n}-V_{n+1}) \boldsymbol{f}(\beta_n,\boldsymbol{r}).
\label{electricFieldBiotPlusStaircaseVContributionEq}
\end{equation}
Here we have defined the vector field 
\begin{equation}
\boldsymbol{f}(\phi',\boldsymbol{r}) := \int_{0}^{\mathscr{R}(\phi')} \frac{(\boldsymbol{r}-\boldsymbol{r}') \times d \rho' \hat{\rho}(\phi')}{|\boldsymbol{r}-\boldsymbol{r}'|^3}
\label{fVectorPrimaryDefEq}
\end{equation}
in Cartesian coordinates with $\hat{\rho}(\phi') = (\cos(\phi'),\sin(\phi'), 0) $. This term is a vector with dimensions of inverse length, and it can be interpreted as a reduced \textit{magnetic field} generated by a unitary current on a straight finite line from a point on $c$ located at $\left( \mathscr{R}(\beta) , \beta  \right)$ (in polar coordinates) to the origin. The integral in Eq.~(\ref{fVectorPrimaryDefEq}) can be evaluated straightforwardly since 
\[
\boldsymbol{f}(\beta,\boldsymbol{r}) = - \hat{n}(\beta) \int_{0}^{\mathscr{R}(\beta)} \frac{\sin\gamma d \rho'}{|\boldsymbol{r}-\boldsymbol{r}'|^2},
\]
with
\[
\hat{n}(\beta) := \frac{\hat{\rho}(\beta) \times \boldsymbol{r}}{\sin\gamma_1},
\]
$\gamma_1$ the angle between $\boldsymbol{r}$ and $\hat{\rho}(\beta)$, and $\gamma$ the angle between $\boldsymbol{r}-\boldsymbol{r}'$ and  $\hat{\rho}(\beta)$. Defining $\tilde{R} = |\boldsymbol{r}-\boldsymbol{r}'| \sin\gamma$ and $s'= - \tilde{R}\cot\gamma$, then $ds'=d\rho'=\tilde{R}\csc^2\gamma d\gamma$, and therefore,
\[
\boldsymbol{f}(\beta,\boldsymbol{r}) = - \frac{\hat{n}(\beta)}{\tilde{R}} \int_{\gamma_1(\beta)}^{\gamma_2(\beta)} \sin\gamma d\gamma.
\]

\begin{figure}[h]
\centering
\includegraphics[width=0.325\textwidth]{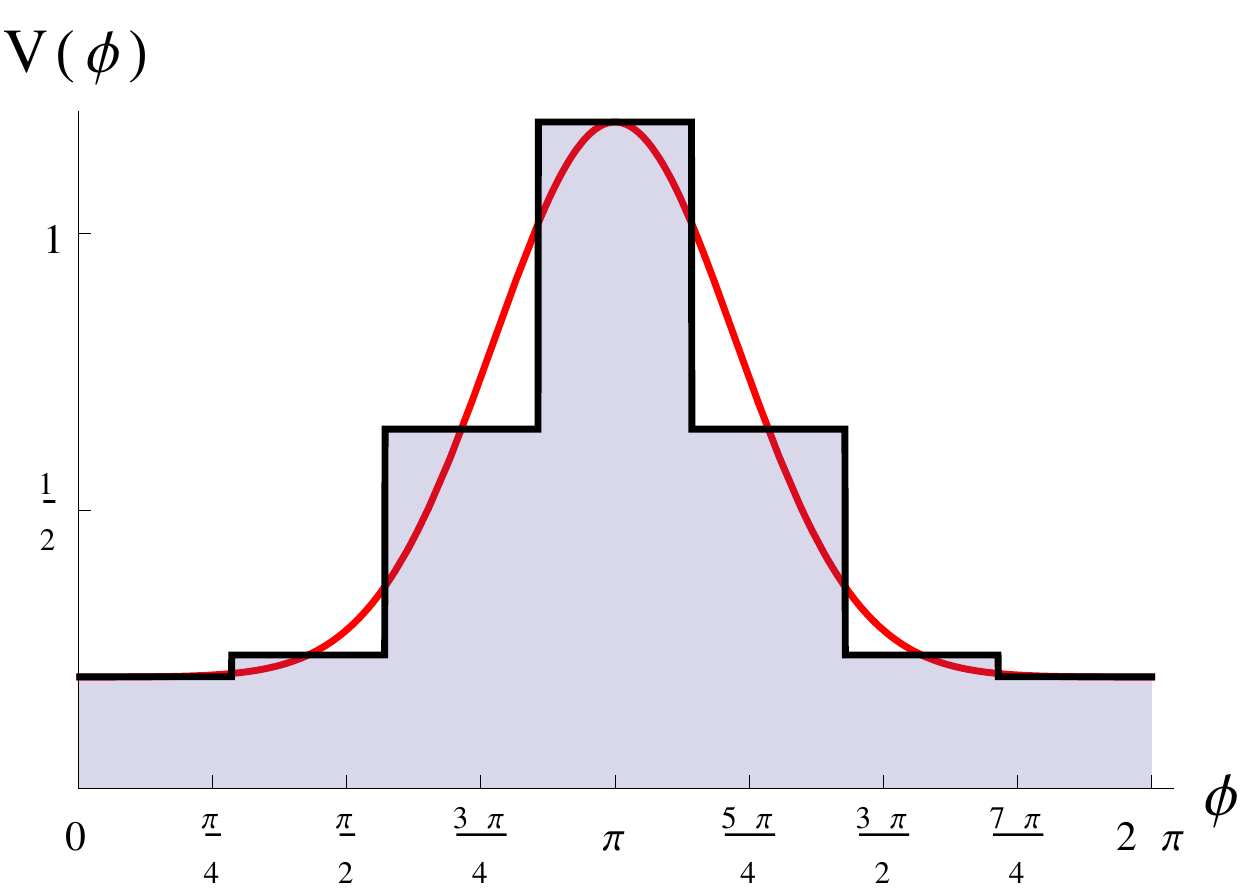}%PotentialN7
\includegraphics[width=0.325\textwidth]{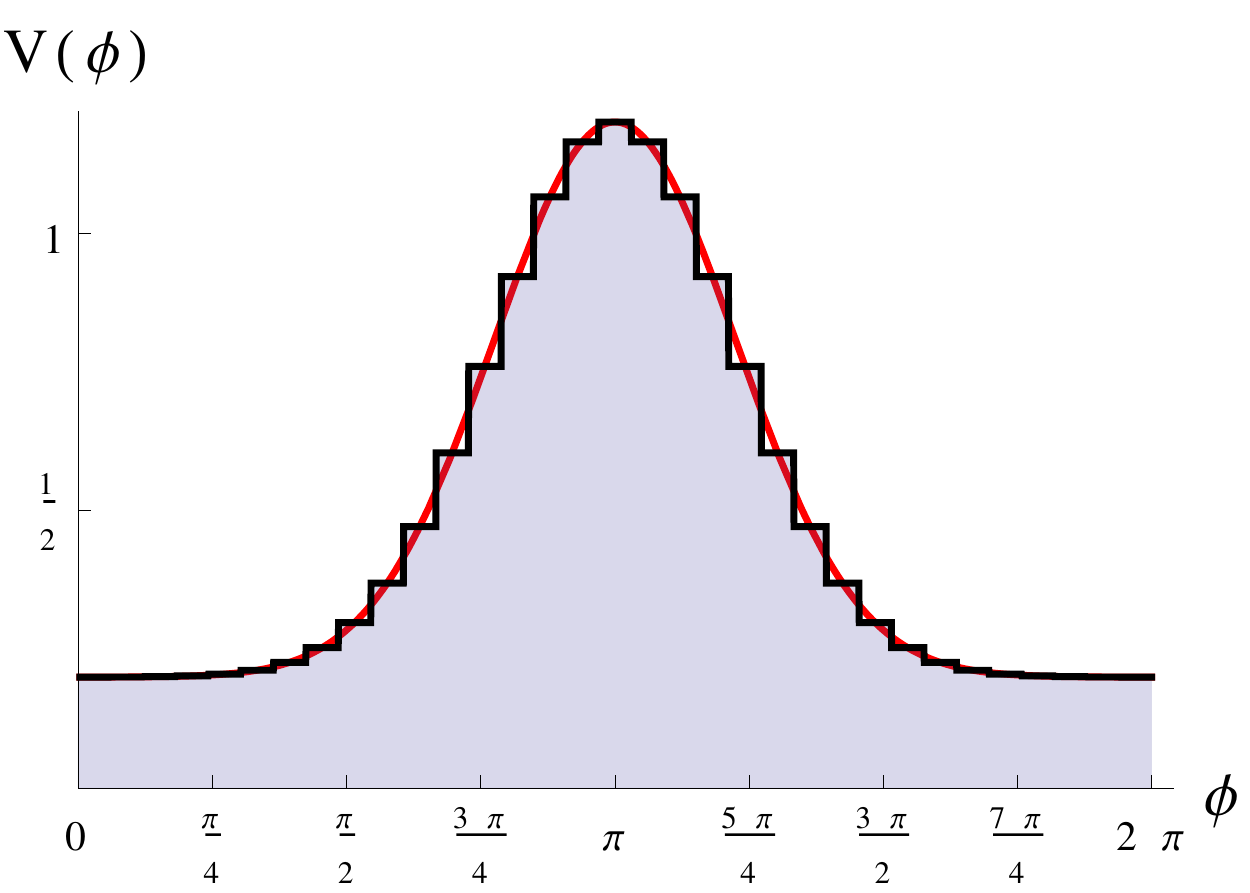}%potentialN33
\includegraphics[width=0.325\textwidth]{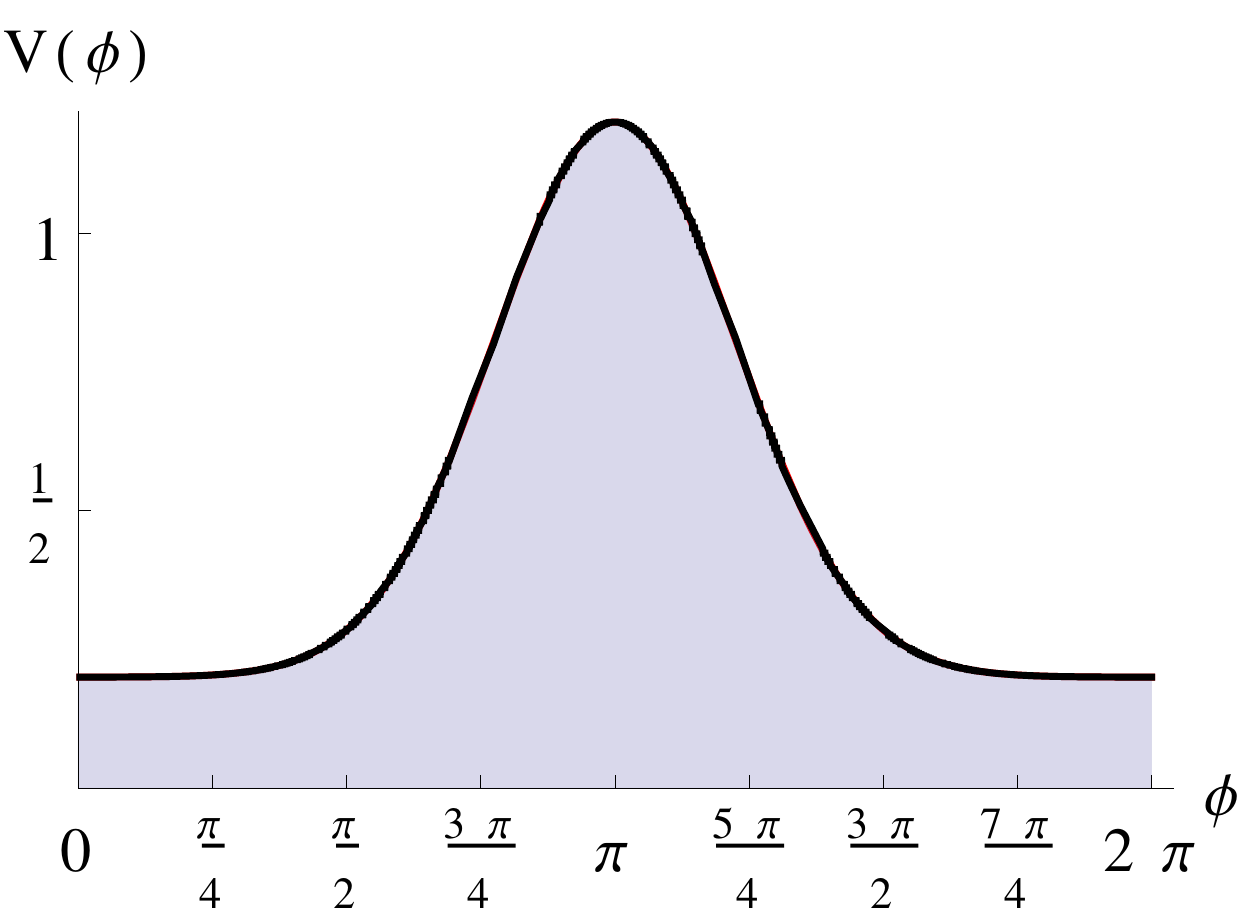}%potentialN444
    \caption{Potential $\mathcal{V}$ (red line) and staircase like potential $V$ (black line) for $N=6, 33$ and 444. $\mathcal{V}$ is selected as smooth periodic function including start and and ending points $\mathcal{V}(0)=\mathcal{V}(2\pi)$. }
\label{discretePotentialLimitFig}
\end{figure}

Since,
\[
\cos\gamma_2(\beta) = \frac{\hat{\rho}(\beta)\cdot \boldsymbol{r} - \mathscr{R}(\beta)}{|\boldsymbol{r}-\boldsymbol{r}'|}
\]
then, $\boldsymbol{f}(\phi',\boldsymbol{r})$ results in 
\begin{equation}
\boldsymbol{f}(\beta,\boldsymbol{r}) = \frac{1}{r} \frac{ \hat{\rho}(\beta) \times \hat{r} }{1-[\hat{\rho}(\beta) \cdot \hat{r}]^2} \left[ \frac{ \hat{\rho}(\beta) \cdot \boldsymbol{r} - \mathscr{R}(\beta) }{|\boldsymbol{r}-\mathscr{R}(\beta)\hat{\rho}(\beta)|} - \hat{\rho}(\beta) \cdot \hat{r}  \right] .
    \label{fVectorWithVectorNotationEq}
\end{equation}

At this point we found an analytic expression for the electric field.
Let us now suppose some potential field distribution in the planar region which fulfils, among other properties, the periodicity conditions.
Indeed, some staircase-like piece-wise distribution $V(\phi)$ of the type 
\[
V(\phi) = \mathcal{V}(\beta_{n-1} + \delta\beta_n /2 )  \hspace{0.5cm}\mbox{\textbf{if}}\hspace{0.5cm}\phi \in (\beta_{n-1},\beta_{n})
\]
would tend to a smooth and continuous potential $\mathcal{V}(\phi)$ as $N\rightarrow\infty$ (see Fig.~(\ref{discretePotentialLimitFig}))
If $\mathcal{V}(\phi)$ is defined as a \textit{fully periodic function} in $\phi\in[0,2\pi]$, that can be expanded in Taylor series as follows
\[
V_{n+1} = \sum_{j=1}^\infty \frac{\delta\beta^j}{j!}\partial_\phi^j \mathcal{V}(\beta_{n}).
\]

Note that not every distribution fulfill this condition. For instance, a smooth linear function $\mathcal{V}(\phi) = \phi$ can be approximated by selecting $V(\phi)$ as an staircase function satisfying periodicity conditions Eq.~(\ref{periodicityConditionsEq}). However, we could not use this linear function $\mathcal{V}(\phi)$ since there would be a discontinuity at some angle, no matter how large becomes $N$. In that situation, $\partial_\phi \mathcal{V}(2\pi)$ is a Dirac delta function, and the Taylor series expansion would diverge. Assuming that $\mathcal{V}(\phi)$ is a suitable distribution, then the expression for a $N\rightarrow\infty$ distribution becomes
\[
\boldsymbol{E}(\boldsymbol{r})  = \frac{1}{2\pi} \oint_{ c } V(\phi') \frac{(\boldsymbol{r}-\boldsymbol{r}') \times d\boldsymbol{r}'}{|\boldsymbol{r}-\boldsymbol{r}'|^3} - \frac{1}{2\pi} \lim_{N \to \infty} \sum_{n=1}^N \partial_\phi V(\beta_{n}) \boldsymbol{f}(\beta_n,\boldsymbol{r}) \delta \beta_n,
\]
which can be replaced to its continuous form 
\[
\boldsymbol{E}(\boldsymbol{r})  = \frac{1}{2\pi} \oint_{ c } V(\phi') \frac{(\boldsymbol{r}-\boldsymbol{r}') \times d\boldsymbol{r}'}{|\boldsymbol{r}-\boldsymbol{r}'|^3} - \frac{1}{2\pi} \int_{0}^{2\pi} \partial_\phi V(\beta) \boldsymbol{f}(\beta,\boldsymbol{r}) d\beta.
\]
Since $V(0)\boldsymbol{f}(0,\boldsymbol{r})=V(2\pi)\boldsymbol{f}(2\pi,\boldsymbol{r})$, we can perform a partial integration on the second term, resulting in the electric field expression given by
\begin{equation}
\boxed{
\boldsymbol{E}(\boldsymbol{r})  = \frac{1}{2\pi} \oint_{ c } V(\phi') \frac{(\boldsymbol{r}-\boldsymbol{r}') \times d\boldsymbol{r}'}{|\boldsymbol{r}-\boldsymbol{r}'|^3} + \langle V, \boldsymbol{f} \rangle,
}
\label{electricFieldBiotSavartWithCorrectionEq}
\end{equation}
with
\[
\langle V, \boldsymbol{f} \rangle := \frac{1}{2\pi} \int_{0}^{2\pi}  V(\beta) \partial_\phi \boldsymbol{f}(\beta,\boldsymbol{r}) d\beta .
\]
The line integral in the right hand of Eq.~(\ref{electricFieldBiotSavartWithCorrectionEq}) can be interpreted as it would be a non-uniform current $V(\phi')\hat{r}'$ (with dimensions of electric field per unit of length) circulating along $c$ and generating electric field in a Biot-Savart like way. The second term $\langle V, \boldsymbol{f} \rangle$ measures the contributions due to a \textit{current} coming form center to the boundary $c$ and weighed on $\mathcal{A}$ with $\partial_\phi V$. As it is expected, the term $\langle V, \boldsymbol{f} \rangle$ vanishes when $V$ is constant. More explicitly, the vector $\boldsymbol{f}$ is 
\[
\boldsymbol{f}(\phi',\boldsymbol{r}) = \mathscr{F}(r,\theta,\phi,\phi') \left[\sin\theta\sin(\phi-\phi')\hat{z} - \cos\theta\hat{\phi}(\phi') \right]
\]
with
\[
\mathscr{F}(r,\theta,\phi,\phi') = \frac{1}{r}\frac{1}{1-\sin^2\theta\cos^2(\phi-\phi')}\left[\frac{r\sin\theta\cos(\phi-\phi') - \mathscr{R}(\phi')}{\sqrt{r^2+\mathscr{R}^2(\phi')-2r\mathscr{R}(\phi')\sin\theta\cos(\phi-\phi')}}-\sin\theta\cos(\phi-\phi')\right]
\]
where $\hat{\phi}(\phi') = -\sin\phi' \hat{x} + cos\phi'\hat{y}$, as usual. Writing the unitary vector of the Cartesian coordinates in terms of the spherical ones and simplifying
\begin{equation}
 \boldsymbol{f}(\phi',\boldsymbol{r}) = \mathscr{F}(r,\theta,\phi,\phi') \left[\sin(\phi-\phi')\hat{\theta}(\boldsymbol{r}) + \cos\theta\cos(\phi-\phi')\hat{\phi}(\boldsymbol{r})\right]
 \label{fVectorSphericalCoordEq}
\end{equation}
therefore $f_r = 0$, which implies that the \textit{radial component of the electric field $E_r(\boldsymbol{r})$ is only the contribution of the Biot-Savart term} since $\langle V, \boldsymbol{f} \rangle$ cannot affect this component. 

\section{Circular region with an arbitrary staircase-like function $V(\phi)$}
In this section we shall find an exact expansion of the electric field generated by a circular region of radius $\mathscr{R}(\phi) = R$ with a staircase-like function $V(\phi)$.  According to Eq.~(\ref{electricFieldBiotPlusStaircaseVContributionEq}) the electric field is
\begin{equation}
\boldsymbol{E}(\boldsymbol{r})  = \pmb{\mathscr{E}}(\boldsymbol{r}) + \frac{\textbf{sgn}(z)}{2\pi}  \sum_{n=1}^N (V_{n}-V_{n+1}) \boldsymbol{f}(\beta_n,\boldsymbol{r})
    \label{electricFieldBiotPlusStaircaseVContributionIIEq}
\end{equation}
where
\[
\pmb{\mathscr{E}}(\boldsymbol{r}) = \frac{\textbf{sgn}(z)}{2\pi} \oint_{ c } V(\phi') \frac{(\boldsymbol{r}-\boldsymbol{r}') \times d\boldsymbol{r}'}{|\boldsymbol{r}-\boldsymbol{r}'|^3}
\]
is the Biot-Savart contribution whose components for $z>0$ are
\[
\mathscr{E}_r(\boldsymbol{r}) = \frac{R^2}{2\pi}\cos\theta \int_{ 0 }^{2\pi}   \frac{V(\phi')}{\mathscr{r}(r,\theta,\phi-\phi')^3} d\phi' ,
\]

\[
\mathscr{E}_\theta(\boldsymbol{r}) = \frac{R r}{2\pi} \int_{ 0 }^{2\pi}   \frac{V(\phi')\cos(\phi-\phi')}{\mathscr{r}(r,\theta,\phi-\phi')^3} d\phi' - \frac{R^2}{2\pi}\sin\theta \int_{ 0 }^{2\pi}   \frac{V(\phi')}{\mathscr{r}(r,\theta,\phi-\phi')^3} d\phi'
\]
and
\[
\mathscr{E}_\phi(\boldsymbol{r}) = -\frac{R r}{2\pi} \cos\theta \int_{ 0 }^{2\pi}   \frac{V(\phi')\sin(\phi-\phi')}{\mathscr{r}(r,\theta,\phi-\phi')^3} d\phi'. 
\]
The magnitude of $\boldsymbol{r}-\boldsymbol{r}'$ is 
\[
\mathscr{r}(r,\theta,\phi-\phi') =\sqrt{R^2+r^2-2Rr\sin\theta\cos(\theta-\theta')}
\]
Since the potential on $\mathcal{A}$ is a staircase-like function then
\[
\mathscr{E}_r(\boldsymbol{r}) = \frac{R^2}{2\pi}\cos\theta \sum_{n=1}^N V_n \int_{ \beta_{n-1} }^{\beta_n} \frac{1}{\mathscr{r}(r,\theta,\phi-\phi')^3} d\phi' .
\]
Analogous expressions can be written for the other two components of the electric field by using 
\[
\int_{ 0 }^{2\pi}   V(\phi') g(\boldsymbol{r},\phi') d\phi' \longrightarrow \sum_{n=1}^N V_n \int_{ \beta_{n-1} }^{\beta_n} g(\boldsymbol{r},\phi') d\phi'.
\]
It is convenient to write the inverse of $\mathscr{r}(r,\theta,\phi-\phi')$ as follows
\[
\frac{1}{\mathscr{r}(r,\theta,\phi-\phi')} = \frac{1}{\sqrt{R^2+r^2}}\frac{1}{\sqrt{1-\xi\cos(\phi-\phi')}} 
\]
with $\xi$ defined as  
\[
\xi(r,\theta) := \frac{2 R r \sin\theta}{R^2+r^2}
\]
in order to use the expansion 
\begin{equation}
\frac{1}{(1-\chi)^{\alpha/2}} = 1 + \frac{\alpha}{2}\chi + \frac{\alpha}{8}(\alpha+2)\chi^2 + \cdots = \sum_{n=0}^{\infty} (-1)^n \binom{-\alpha/2}{n} \chi^n 
\label{binomialTheoremSeriesExpansionEq}    
\end{equation}
to write
\begin{equation}
\frac{\cos^m(\phi-\phi')}{\mathtt{r}(r,\theta,\phi-\phi')^\alpha} = \frac{1}{\left(\sqrt{R^2+r^2}\right)^\alpha} \sum_{n=0}^{\infty} (-1)^n \binom{-\alpha/2}{n} \xi^n \left\{ \zeta_{n+m} + \frac{2}{2^{n+m}} \sum_{k=0}^{\floor*{(n+m-1)/2}} \binom{n+m}{k} \cos[\nu_{nmk}(\phi-\phi')] \right\}
\label{cosBetaOverREq}
\end{equation}
for any positive integer $m$ (including zero), $\nu_{nmk}=n+m-2k$, $\floor*{z}$ the floor function and
\[
\zeta_n := \frac{1}{2^n} \binom{n}{n/2} \hspace{0.5cm}\mbox{\textbf{if}}\hspace{0.5cm} n \in 2\mathbb{N}^0 \hspace{0.5cm}\mbox{\textbf{else}}\hspace{0.5cm} 0.
\]

\begin{figure}[h]
  \begin{minipage}[b]{0.33\linewidth}
    
   \includegraphics[width=1.0\textwidth]{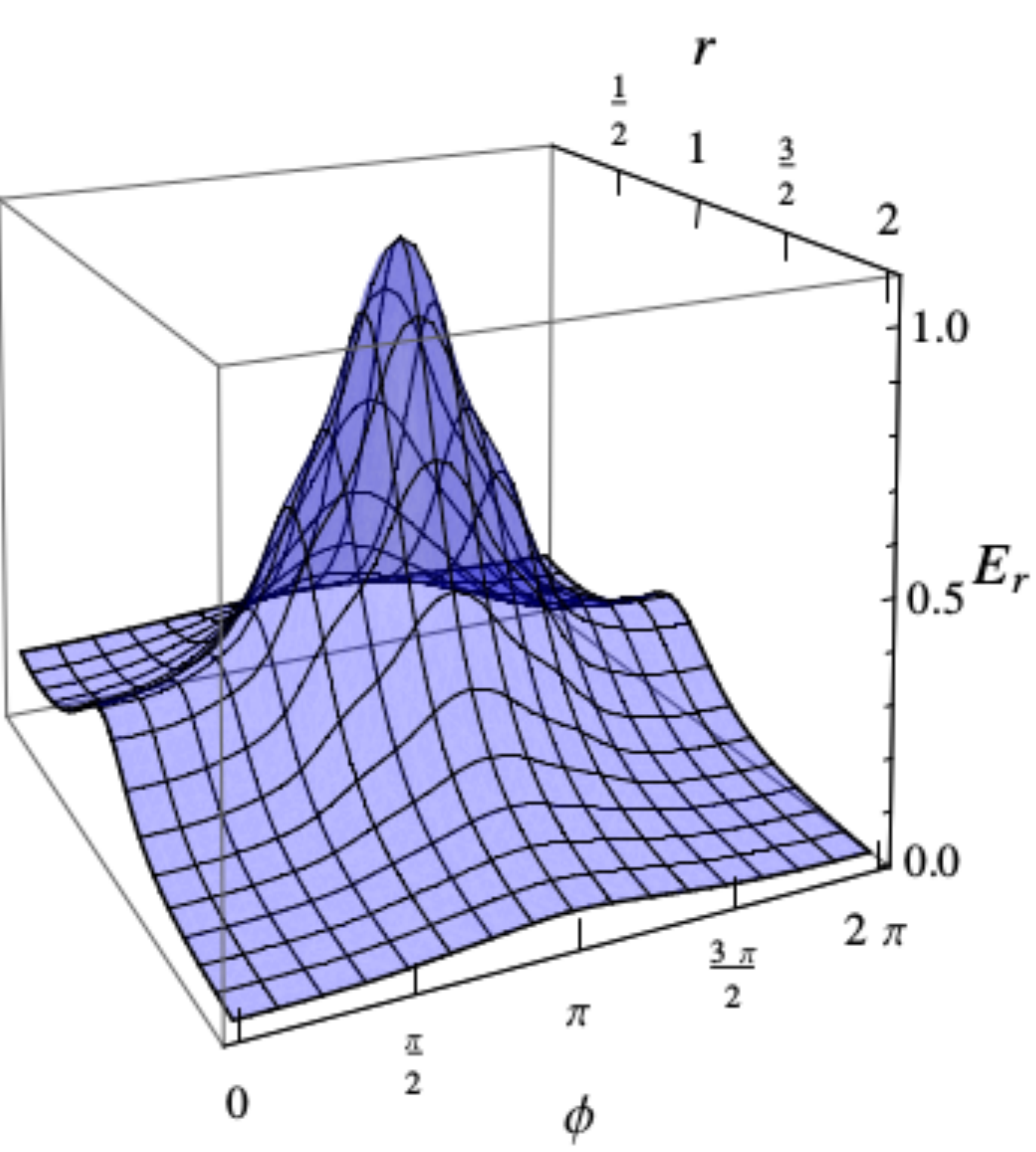}
   \caption*{(a)}%ErSurfaceN7.pdf
    
  \end{minipage} 
  \begin{minipage}[b]{0.33\linewidth}
    
    \includegraphics[width=1.0\textwidth]{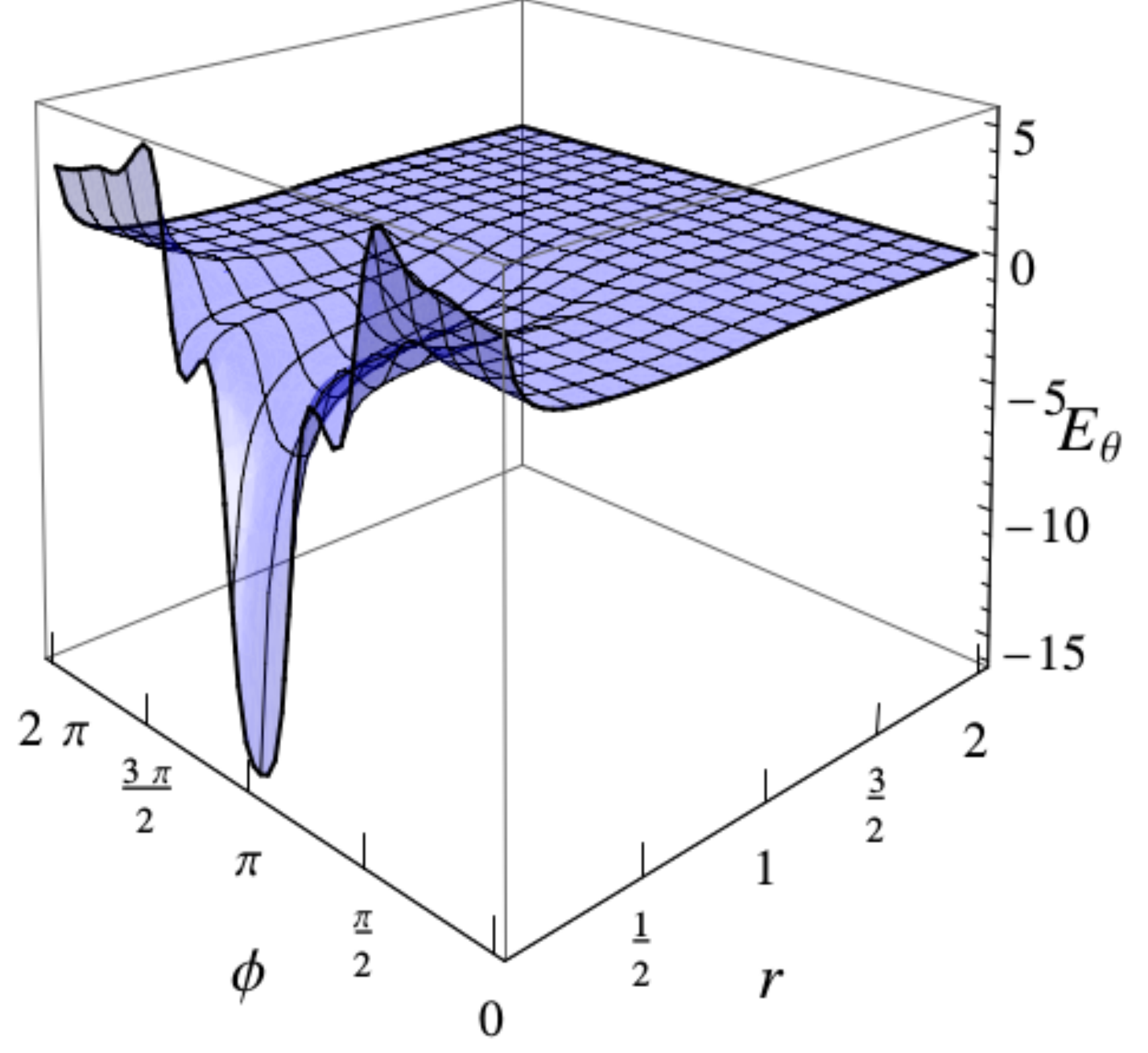}
    \caption*{(b) } %EThetaSurfaceN7.pdf
  
  \end{minipage} 
  \begin{minipage}[b]{0.33\linewidth}
    
    \includegraphics[width=1.0\textwidth]{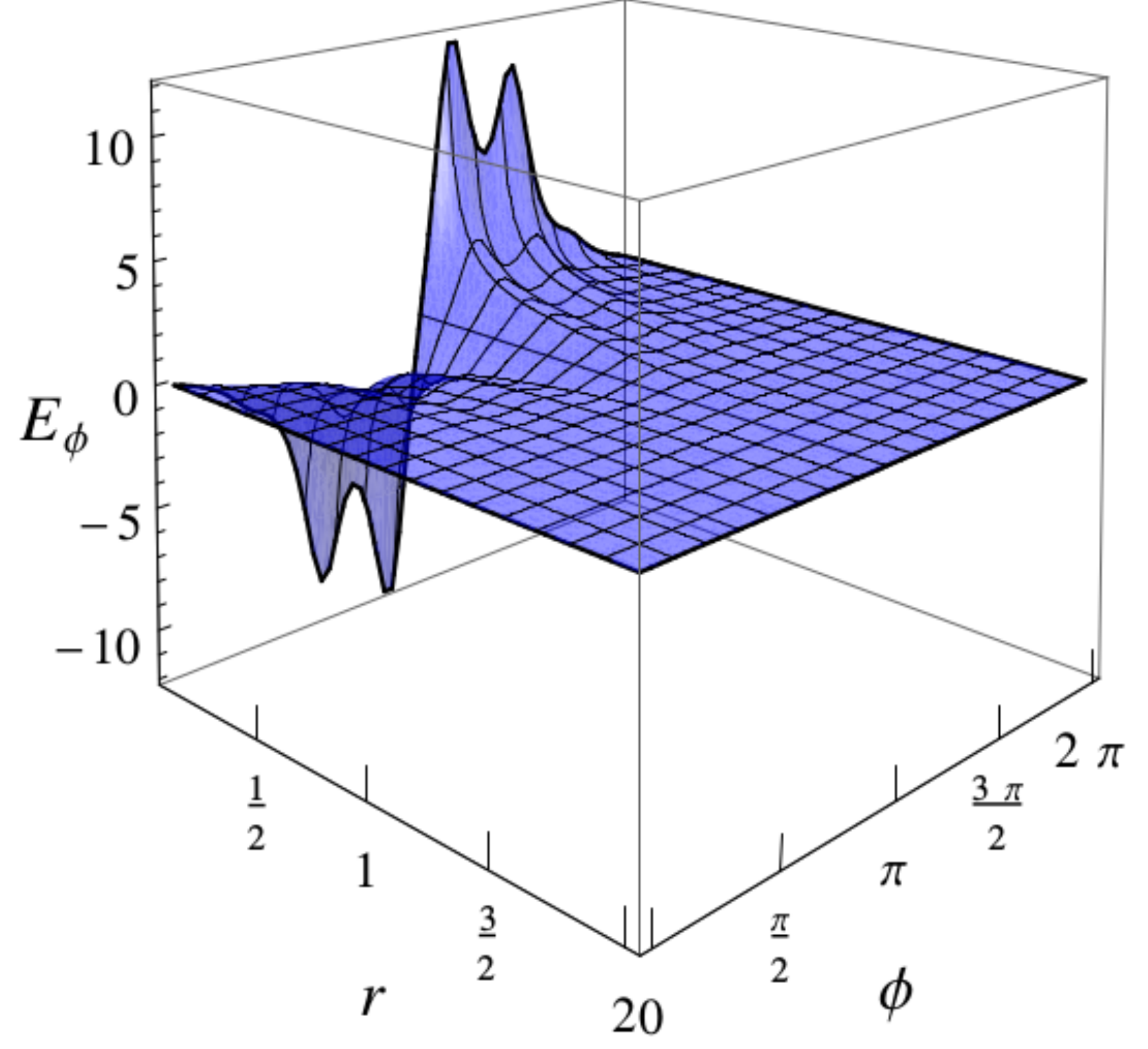}
    \caption*{(c) } %EphiSurfaceN7.pdf
  
  \end{minipage}
  \caption{Components of the electric field. In all the plots we have set $\theta = 2\pi/5$ and using the analytic formula in Eq.~(\ref{EVectorFieldStairCaseVExpansionEq}). (a), (b) and (c) corresponds to the components $\boldsymbol{E}(\boldsymbol{r})$ keeping $V(\phi)$ as the one shown in Fig.~\ref{discretePotentialLimitFig}-(left).}
   \label{EComponentsSurfFig} 
\end{figure}

We may use the previous expansions to evaluate the integrals of Biot-Savart contribution. For instance, the following integral  
\begin{equation}
    \sum_{n=1}^N V_n \int_{ \beta_{n-1} }^{\beta_n} \frac{\cos^m (\phi-\phi')}{\mathscr{r}(r,\theta,\phi-\phi')^3} d\phi'
    \label{cosIntegralDefEq}
\end{equation}
is required to compute $\mathscr{E}_r(\boldsymbol{r})$ and $\mathscr{E}_\theta(\boldsymbol{r})$ where $m$ should be set as 0 or 1. Then using the series expansion in Eq.~(\ref{cosBetaOverREq}) we find 
\[
\begin{split}
\sum_{n=1}^N V_n \int_{ \beta_{n-1} }^{\beta_n} \frac{\cos^m (\phi-\phi')}{\mathscr{r}(r,\theta,\phi-\phi')^3} d\phi' & =  \frac{1}{\left(R^2+r^2\right)^{3/2}} \sum_{n=0}^{\infty} (-1)^s \binom{-3/2}{s} \xi^s \left\{  \zeta_{s+m} \sum_{n=1}^{N-1} V_n(\beta_{n} - \beta_{n-1}) + \right.  \\ & \left. \frac{2}{2^{s+m}} \sum_{k=0}^{\floor*{(s+m-1)/2}} \binom{s+m}{k}  \sum_{n=1}^N \frac{\sin[\nu_{smk}(\phi-\beta_{n-1})] -  \sin[\nu_{smk}(\phi-\beta_{n})]}{\nu_{smk}}  \right\}
\end{split}
\]
Now
\[
\sum_{n=1}^N V_n(\beta_n - \beta_{n-1}) = \sum_{n=1}^N V_n \beta_n - \sum_{n=0}^{N-1} V_{n+1}\beta_n  
\]
since $\beta_N = 2\pi$ and $\beta_0 = 0$ then 
\[
\sum_{n=1}^N V_n(\beta_n - \beta_{n-1}) = 2\pi V_N + \sum_{n=1}^{N-1} (V_n - V_{n+1})\beta_n   .
\]

\begin{figure}[h] 
  \begin{minipage}[b]{0.33\linewidth}
    
   \includegraphics[width=1.0\textwidth]{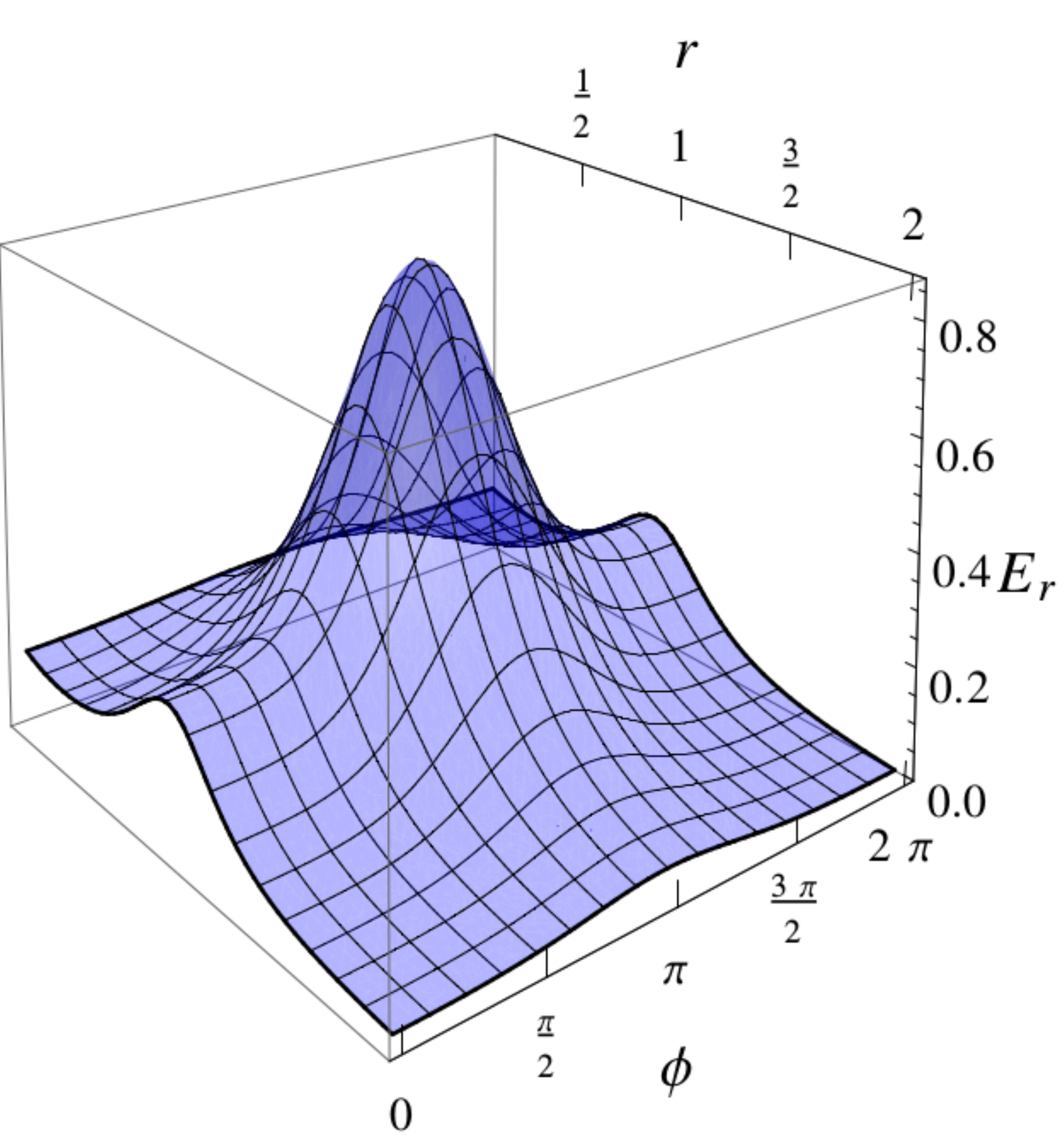}
   \caption*{(a)}%ErSurfaceN33.pdf
    
  \end{minipage} 
  \begin{minipage}[b]{0.33\linewidth}
    
    \includegraphics[width=1.0\textwidth]{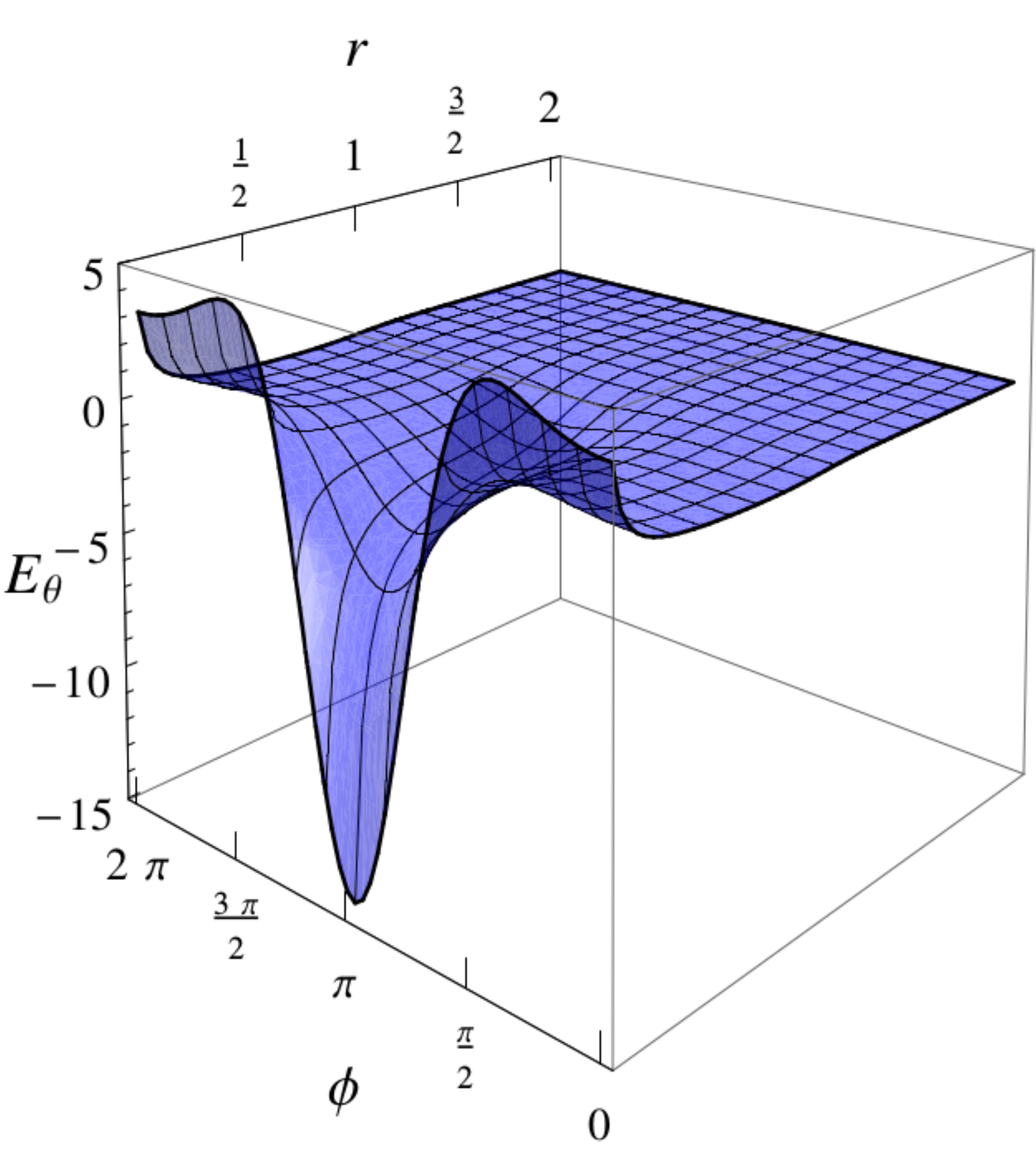}
    \caption*{(b) } %EThetaSurfaceN33.pdf
  
  \end{minipage} 
  \begin{minipage}[b]{0.33\linewidth}
    
    \includegraphics[width=1.0\textwidth]{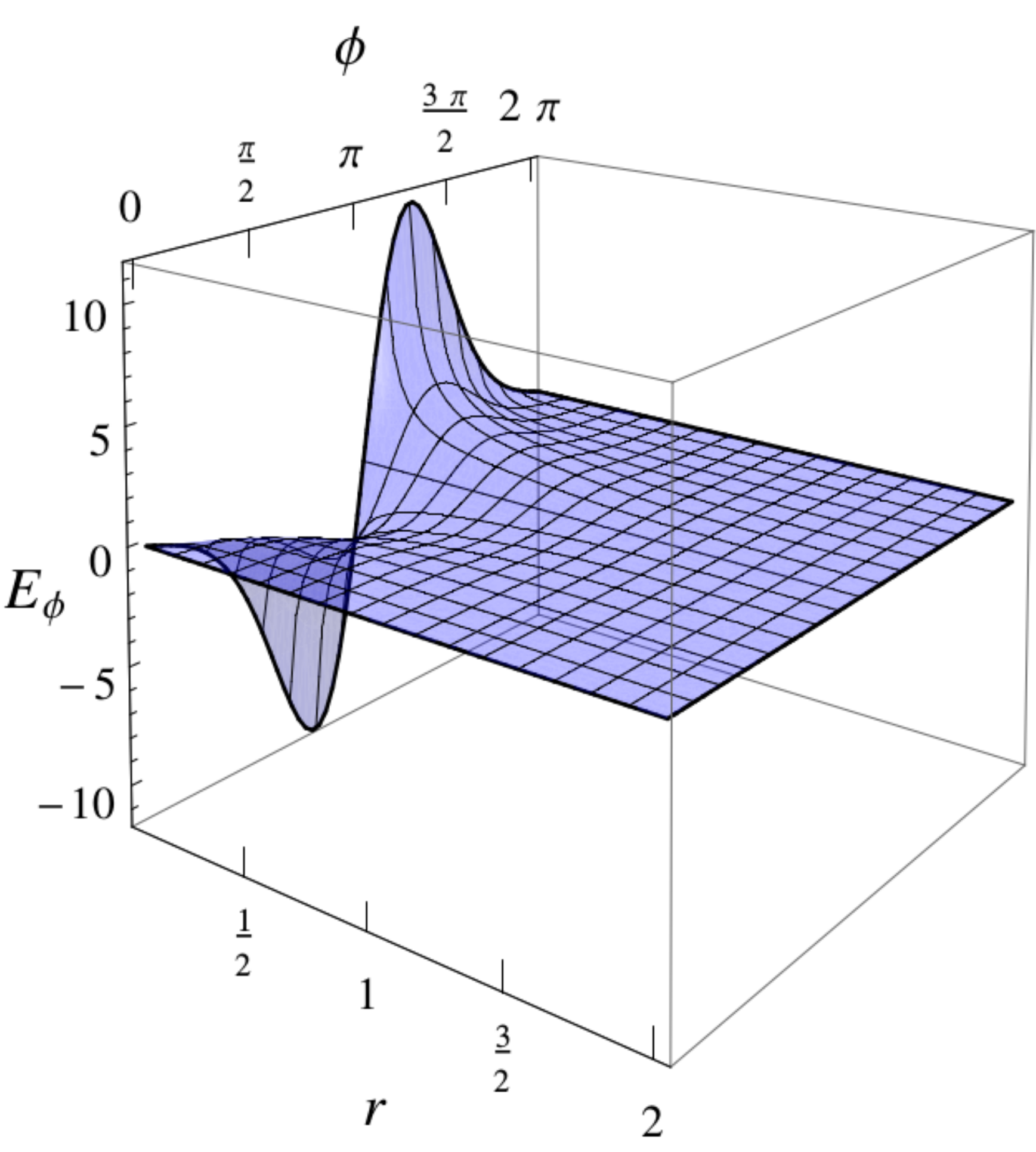}
    \caption*{(c) } %EphiSurfaceN33.pdf
  
  \end{minipage}
  \caption{Components of the electric field. In all the plots we have set $\theta = 2\pi/5$ and using the analytic formula in Eq.~(\ref{EVectorFieldStairCaseVExpansionEq}). (a), (b) and (c) corresponds to the components $\boldsymbol{E}(\boldsymbol{r})$ keeping $V(\phi)$ as the one shown in Fig.~\ref{discretePotentialLimitFig}-(right).}
  
  \label{EComponentsSurfIIFig} 
\end{figure}

Similarly
\[
\sum_{n=1}^N \left\{\sin[\nu(\phi-\beta_{n})] -  \sin[\nu(\phi-\beta_{n-1})]\right\} = \sum_{n=1}^N (V_n - V_{n+1}) \sin[\nu(\phi-\beta_{n})]  \hspace{0.5cm}\forall\hspace{0.5cm}  \nu\in \mathbb{Z}^0
\]
where we have used the periodicity conditions of Eq.~(\ref{periodicityConditionsEq}), hence
\[
\begin{split}
\sum_{n=1}^N V_n \int_{ \beta_{n-1} }^{\beta_n} \frac{\cos^m (\phi-\phi')}{\mathscr{r}(r,\theta,\phi-\phi')^3} d\phi' & =  \frac{1}{\left(R^2+r^2\right)^{3/2}} \sum_{n=0}^{\infty} (-1)^s \binom{-3/2}{s} \xi^s \left\{  \zeta_{s+m} \left[\sum_{n=1}^{N-1} (V_n - V_{n+1})\beta_n + 2\pi V_N \right] - \right.  \\ & \left. \frac{2}{2^{s+m}} \sum_{k=0}^{\floor*{(s+m-1)/2}} \binom{s+m}{k}  \frac{1}{\nu_{smk}} \sum_{n=1}^N (V_n - V_{n+1}) \sin[\nu_{smk}(\phi-\beta_{n})]   \right\}   .
\end{split}  
\]
Note that if the potential would be uniform $V_n = V_1 \hspace{0.25cm} \forall \hspace{0.25cm} n \in \{1,2,\ldots,N\}$, then the previous integral (divided by $V_1$) would take the form
\begin{equation}
\int_{0}^{2\pi} \frac{\cos^m (\phi-\phi')}{\mathscr{r}(r,\theta,\phi-\phi')^3} d\phi' = \frac{1}{\left(R^2+r^2\right)^{3/2}} \sum_{n=0}^{\infty} (-1)^s \binom{-3/2}{s} \xi^s \zeta_{s+m} (2\pi) 
\label{integralExpansionForUniformRingEq}
\end{equation}
therefore
\begin{equation}
\begin{split}
\sum_{n=1}^N V_n \int_{ \beta_{n-1} }^{\beta_n} \frac{\cos^m (\phi-\phi')}{\mathscr{r}(r,\theta,\phi-\phi')^3} d\phi' & =  V_1 \int_{0}^{2\pi} \frac{\cos^m (\phi-\phi')}{\mathscr{r}(r,\theta,\phi-\phi')^3} d\phi' + \\ & \left. \frac{1}{\left(R^2+r^2\right)^{3/2}} \sum_{n=0}^{\infty} (-1)^s \binom{-3/2}{s} \xi^s \sum_{n=1}^N (V_n - V_{n+1}) \tau_{m,n}^s(\phi)   \right\}
\end{split}  
\label{cosIntegralExpansionEq}
\end{equation}
where we have defined
\[
\tau_{m,n}^s (\phi) :=  \zeta_{s+m}\beta_n - \frac{2}{2^{s+m}} \sum_{k=0}^{\floor*{(s+m-1)/2}} \binom{s+m}{k}  \frac{1}{\nu_{smk}} \sin[\nu_{smk}(\phi - \beta_{n})].
\]
Another integral required to compute $\mathscr{E}_\phi(\boldsymbol{r})$ is
\begin{equation}
\sum_{n=1}^N V_n \int_{ \beta_{n-1} }^{\beta_n} \frac{\sin (\phi-\phi')}{\mathscr{r}(r,\theta,\phi-\phi')^3} d\phi'  .
\label{sinIntegralDefEq}
\end{equation}
We may employ the same strategy used in Eq.~(\ref{cosIntegralDefEq}) to evaluate Eq.~(\ref{sinIntegralDefEq}) by setting $m=0$, using
\[
\int_{\beta_{n-1}}^{\beta_n} \cos[\nu(\phi-\phi')] \sin(\phi-\phi')d\phi' = \frac{1}{2}\sum_{\sigma\in\left\{-1,1\right\}} \frac{\sigma}{\nu+\sigma}\left\{ \cos[(\nu+\sigma)(\beta_n-\phi)] - \cos[(\nu+\sigma)(\beta_{n-1}-\phi)] \right\}
\]
if $|\nu| \neq 1$, otherwise
\[
\int_{\beta_{n-1}}^{\beta_n} \cos[(\phi-\phi')] \sin(\phi-\phi')d\phi' = \frac{1}{4}\left\{\cos[2(\phi-\beta_{n})]-\cos[2(\phi-\beta_{n-1})]\right\} .
\]
The result is the following 
\begin{equation}
\sum_{n=1}^N V_n \int_{ \beta_{n-1} }^{\beta_n} \frac{\sin (\phi-\phi')}{\mathscr{r}(r,\theta,\phi-\phi')^3} d\phi' = \frac{1}{\left(R^2+r^2\right)^{3/2}} \sum_{n=0}^{\infty} (-1)^s \binom{-3/2}{s} \xi^s \sum_{n=1}^N (V_n - V_{n+1}) \tau^{(s)}(\phi - \beta_{n})
\label{sinIntegralExpansionEq}
\end{equation}
where we used the periodicity condition given by Eq.~(\ref{periodicityConditionsEq}) and $\tau^{(s)}(\phi)$ is defined as follows
\[
\tau^{(s)}(\phi) =  \zeta_{s}\cos(\phi) + \frac{1}{2^{s}} \sum_{k=0}^{\floor*{(s-1)/2}} \binom{s}{k} \left\{ \sum_{\sigma\in\left\{-1,1\right\}}  \left(\frac{\sigma}{\nu_{sk}+\sigma}\right) \cos[(\nu_{sk}+\sigma)\phi] \hspace{0.25cm}\mbox{\textbf{if}}\hspace{0.25cm} |s-2k| \neq 1 \hspace{0.25cm}\mbox{\textbf{else}}\hspace{0.25cm} \frac{\cos(2\phi)}{2} \right\}.
\]

The radial component of the Biot-Savart contribution takes the form
 \[
 \mathscr{E}_r(\boldsymbol{r}) = \frac{R^2 \cos\theta}{2\pi} \left\{ V_1\int_{0}^{2\pi} \frac{d\phi'}{\mathscr{r}(r,\theta,\phi-\phi')^3} + \frac{1}{\left(R^2+r^2\right)^{3/2}} \sum_{n=0}^{\infty} (-1)^s \binom{-3/2}{s} \xi^s \sum_{n=1}^N (V_n - V_{n+1}) \tau_{0,n}^s(\phi) \right\} .
 \]

Note that the integral term
\[
\left( \mathscr{E}_r(\boldsymbol{r}) \right)_{unif.} = \frac{R^2 \cos\theta}{2\pi} V_1\int_{0}^{2\pi} \frac{d\phi'}{\mathscr{r}(r,\theta,\phi-\phi')^3}
\]
is the radial component of a electric field generated by a circular region with fixed potential $V_1$ inside it and zero at the rest of the $xy$-plane. Similarly, 
\[
 \mathscr{E}_{\theta}(\boldsymbol{r}) =  \left( \mathscr{E}_{\theta}(\boldsymbol{r} ) \right)_{unif.} +\frac{1}{2\pi} \frac{R}{\left(R^2+r^2\right)^{3/2}} \sum_{n=0}^{\infty} (-1)^s \binom{-3/2}{s} \xi^s \sum_{n=1}^N (V_n - V_{n+1})\left[ r \tau_{1,n}^s(\phi) - R\sin\theta \tau_{0,n}^s(\phi) \right]
 \]
where
\[
\left( \mathscr{E}_{\theta}(\boldsymbol{r}) \right)_{unif.} = \frac{R r V_1}{2\pi} \int_{ 0 }^{2\pi}   \frac{\cos(\phi-\phi')}{\mathscr{r}(r,\theta,\phi-\phi')^3} d\phi' - \frac{R^2 V_1}{2\pi}\sin\theta \int_{ 0 }^{2\pi}   \frac{1}{\mathscr{r}(r,\theta,\phi-\phi')^3} d\phi'
\]
and 
\[
 \mathscr{E}_{\phi}(\boldsymbol{r}) = \left( \mathscr{E}_{\phi}(\boldsymbol{r}) \right)_{unif.} -\frac{Rr\cos\theta}{2\pi} \frac{1}{\left(R^2+r^2\right)^{3/2}} \sum_{n=0}^{\infty} (-1)^s \binom{-3/2}{s} \xi^s \sum_{n=1}^N (V_n - V_{n+1}) \tau^{(s)}(\phi-\beta_n)
\]
with $\left( \mathscr{E}_{\phi}(\boldsymbol{r}) \right)_{unif.} = 0$ because the circular region at constant potential is axially symmetric. We may use the fact that $\pmb{\mathscr{E}}(\boldsymbol{r})_{unif.}$ is closely related with magnetic field $\boldsymbol{B}(\boldsymbol{r})_{ring}$ of a circular loop carrying a uniform current $i_o$ by a proportional constant equal to $\mu_o i_o / 2 V_1$ where $\mu_o$ is the magnetic permeability. In general, the magnetic field can be obtained via the vector potential, in our case we may also associate a vector potential $\boldsymbol{A}(\boldsymbol{r})$ such that $\pmb{\mathscr{E}}(\boldsymbol{r})_{unif.} = \textbf{sgn}(z) \mbox{curl} \boldsymbol{A}(\boldsymbol{r})$ where     
\[
 \boldsymbol{A}(\boldsymbol{r}) = \frac{V_1}{2\pi} \frac{4  R}{R^2+r^2+2Rr\sin\theta} \left[ \frac{(2-\gamma^2)K(\gamma^2)-2\b{E}(\gamma^2)}{\gamma^2} \right] \hat{\phi}(\boldsymbol{r})
\]
with 
\[
\gamma^2 = \frac{4 R r \sin\theta}{R^2+r^2+2Rr\sin\theta}
\]
and $K(\gamma^2)$, $\b{E}(\gamma^2)$ the complete elliptic integrals of the first and second kind respectively \cite{jackson1999classical,abramowitz1965handbook}. The bar under letter E is used to avoid confusions between the electric field and the elliptic integral. 

\begin{figure}[H]  
  \begin{minipage}[b]{0.45\linewidth}
    
   \includegraphics[width=1\textwidth]{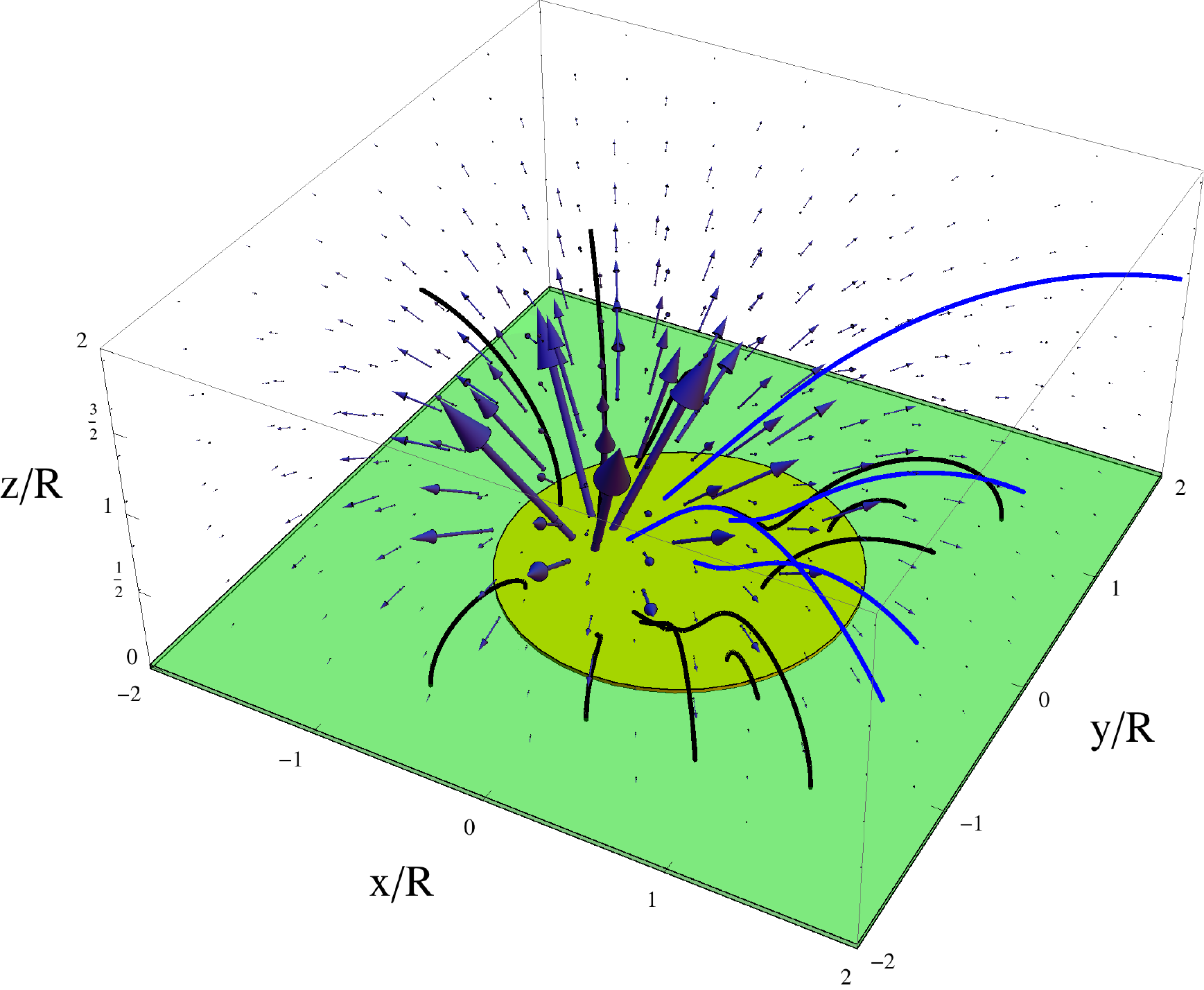}
   \caption*{(a)}%EVectorFieldPlotN33.pdf
    
  \end{minipage} 
  \begin{minipage}[b]{0.5\linewidth}
    
    \includegraphics[width=1\textwidth]{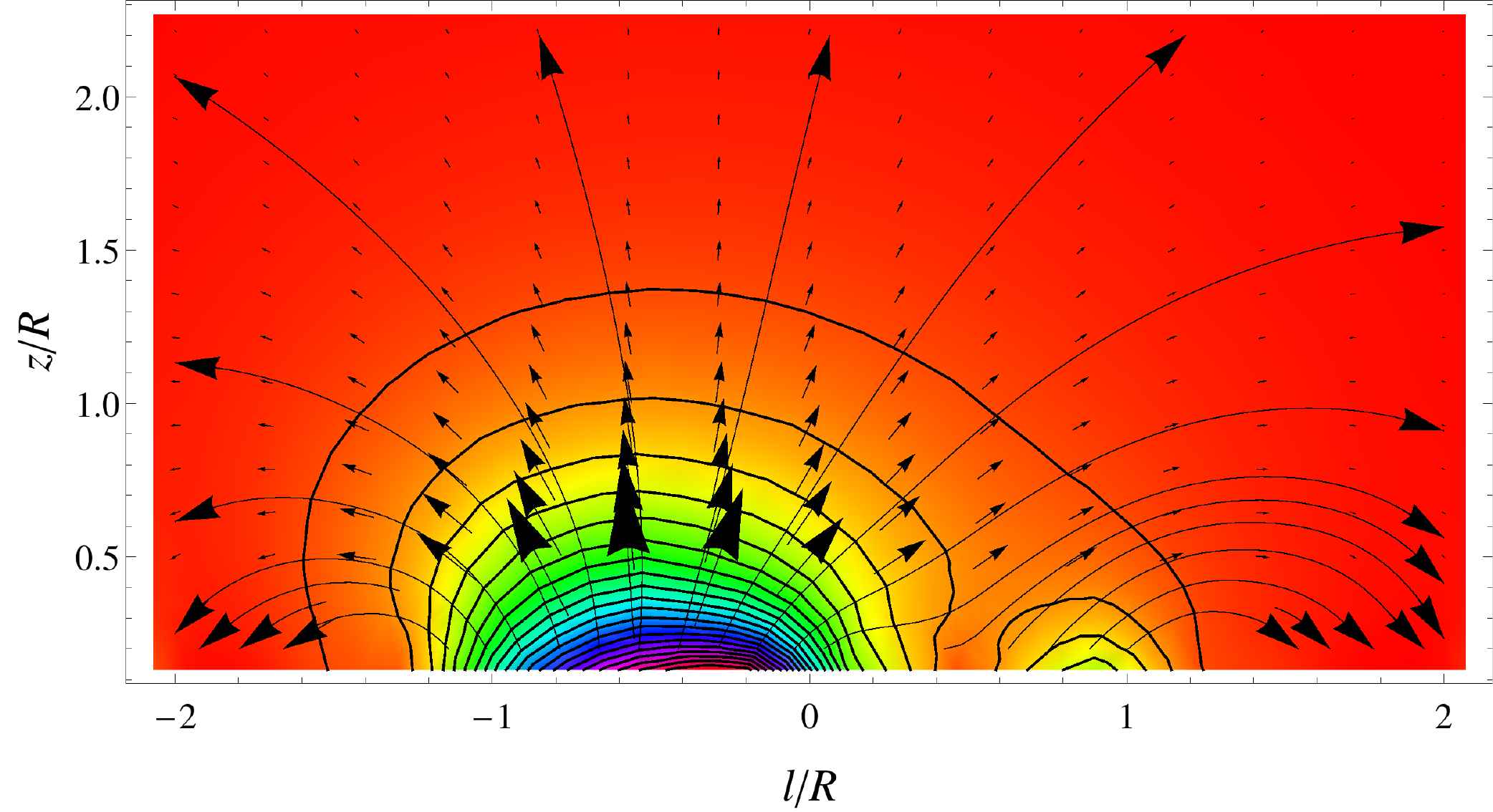}
    \caption*{(b) $\phi=0$} %vectorFieldPhiZeroN33.pdf
  
  \end{minipage} 
  \begin{minipage}[b]{0.5\linewidth}
    
    \includegraphics[width=1\textwidth]{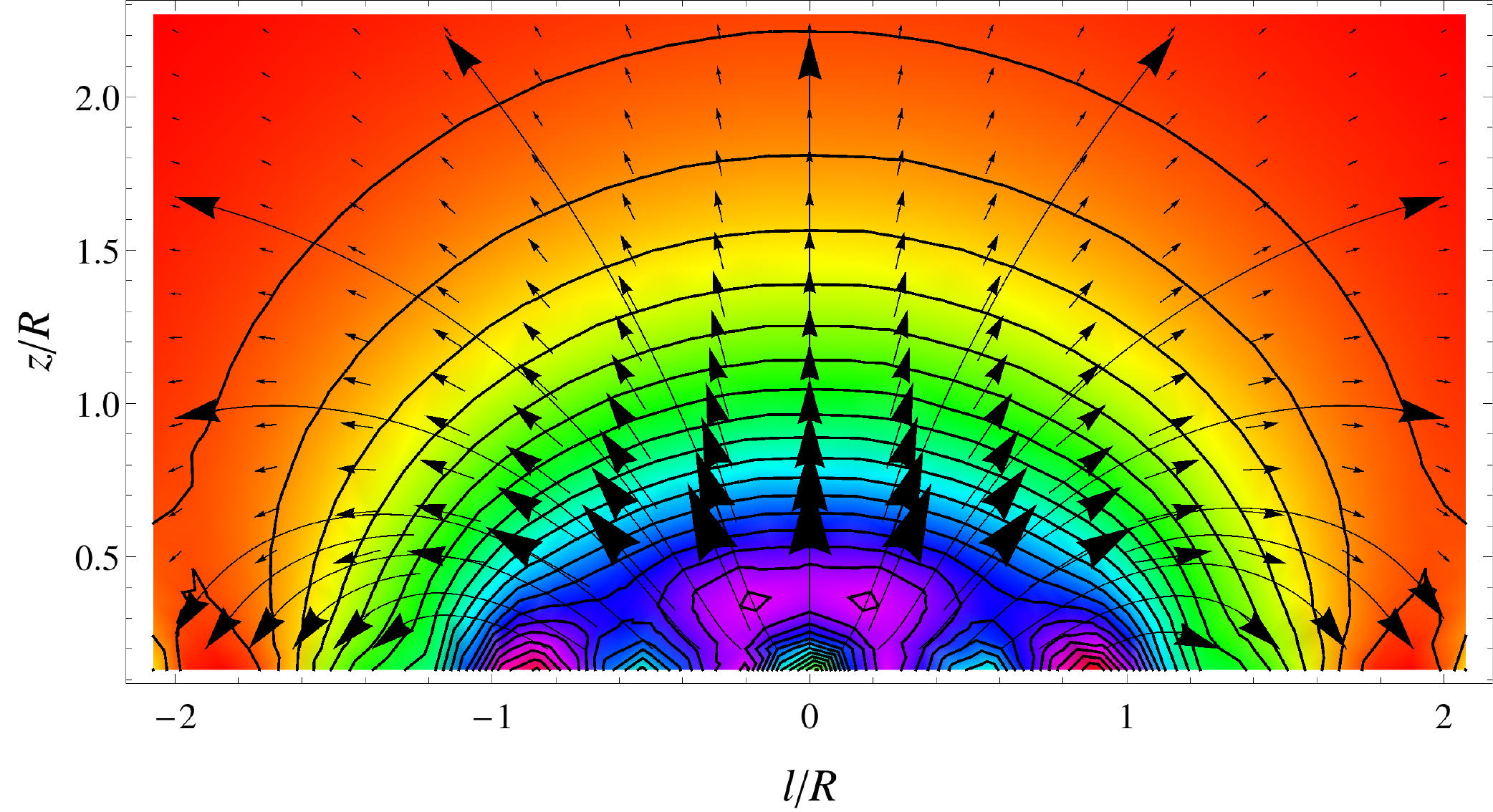}
    \caption*{(c) $\phi=\pi/2$} %vectorFieldPhi05PiN33.pdf
  
  \end{minipage}
  \hfill
  \begin{minipage}[b]{0.5\linewidth}
    \includegraphics[width=1\textwidth]{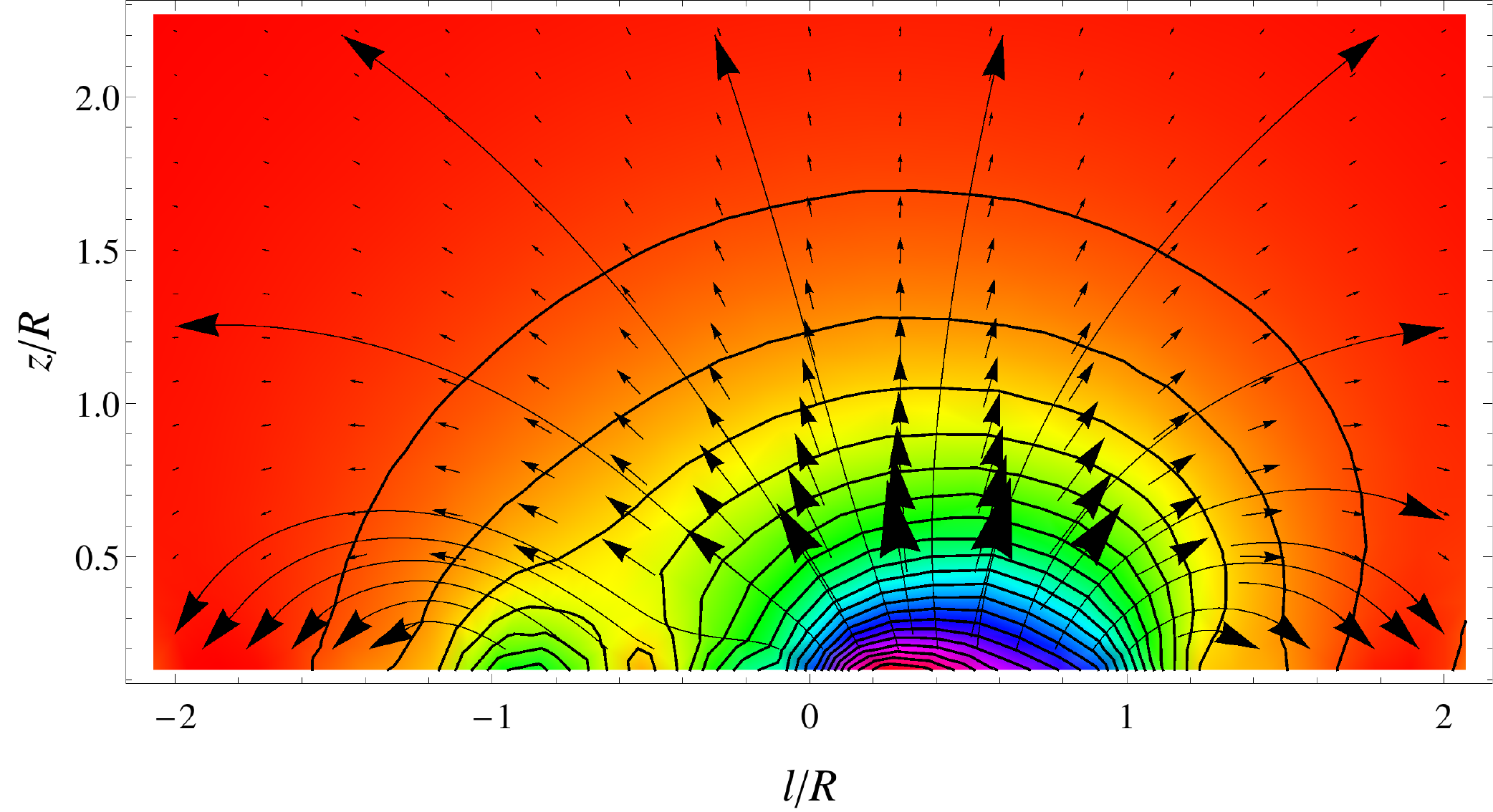}
    \caption*{(d) $\phi=3\pi/4$} %vectorFieldPhi3Pi4N33.pdf
    
  \end{minipage}
  
  \caption{ Electric field. (a) Vector field and few stream lines computed with Eq.~(\ref{EVectorFieldStairCaseVExpansionEq}). Computations were performed for a Gaussian potential $\mathcal{V}$ defined in Eq.~(\ref{gaussianContinousVEq}) and setting $N=33$. (b), (c) and (d) are projections of the $\boldsymbol{E}(\boldsymbol{r})$ on the plane $\phi=0, \pi/2$ and $3\pi/4$ respectively.  }
  
  \label{vectorFieldEPhiConstFig}
  
\end{figure}

The application of the rotational on the vector potential can be demanding algebraically but it is a standard problem that was already solved \cite{garrett1963calculation,simpson2001simple}. An alternative is to use the series expansions of Eq.~(\ref{integralExpansionForUniformRingEq}) to evaluate $\left( \mathscr{E}_{r}(\boldsymbol{r}) \right)_{unif.}$ and $\left( \mathscr{E}_{\theta}(\boldsymbol{r}) \right)_{unif.}$ by setting $m=1$ or $m=0$. The components of $\left(\pmb{\mathscr{E}}(\boldsymbol{r})\right)_{unif.}$ for $z > 0$ are  
\[
\left( \mathscr{E}_{r}(\boldsymbol{r}) \right)_{unif.} = \frac{R^2 V_1}{2\pi} \frac{4\cos\theta}{\mathscr{r}_{-}(r,\theta) \mathscr{r}_{+}(r,\theta)^2} \b{E}\left[ -\frac{4rR\sin\theta}{ \mathscr{r}_{-}(r,\theta)^2 }\right]
\]
and  
\[
\left( \mathscr{E}_{\theta}(\boldsymbol{r}) \right)_{unif.} = \frac{R r V_1}{2\pi} \frac{2\csc\theta}{r R \mathscr{r}_{-}^2 \mathscr{r}_{+}} \left[  (r^2+R^2)\b{E}\left( \frac{4rR\sin\theta}{ \mathscr{r}_{+}^2 }\right) - \mathscr{r}_{-}^2 K\left( \frac{4rR\sin\theta}{ \mathscr{r}_{+}^2 }\right)\right] 
- \frac{R^2 V_1}{2\pi}\frac{4\sin\theta}{\mathscr{r}_{-}\mathscr{r}_{+}^2} \b{E}\left( - \frac{4rR\sin\theta}{ \mathscr{r}_{-}^2 }\right)
\]
where we have defined
\[
\mathscr{r}_{\pm}(r,\theta) = \sqrt{r^2 + R^2 \pm 2rR\sin\theta}.
\]

The Biot-Savart contribution can be written as follows
\[
\boldsymbol{\mathscr{E}}(\boldsymbol{r}) = \left( \boldsymbol{\mathscr{E}}(\boldsymbol{r}) \right)_{unif.} + \frac{1}{2\pi} \frac{R}{\left(R^2+r^2\right)^{3/2}} \sum_{n=1}^N (V_n - V_{n+1}) \sum_{n=0}^{\infty} (-1)^s \binom{-3/2}{s} \xi^s \boldsymbol{L}_n^{(s)}(\boldsymbol{r})
\]
where $\boldsymbol{L}_n^{(s)}(\boldsymbol{r})$ is a vector with length units given by 
\[
\boldsymbol{L}_n^{(s)}(\beta_n,\boldsymbol{r}) = R\cos\theta\tau_{0,n}^s(\phi) \hat{r}(\boldsymbol{r}) + \left[ r \tau_{1,n}^s(\phi) - R\sin\theta \tau_{0,n}^s(\phi) \right]\hat{\theta}(\boldsymbol{r}) - r\cos\theta \tau^{(s)}(\phi-\beta_n)\hat{\phi}(\boldsymbol{r}) 
\]
in spherical coordinates. We can replace the Biot-Savart contribution in Eq.~(\ref{electricFieldBiotPlusStaircaseVContributionIIEq}) to write the total electric field in this compact way
\begin{equation}
\boxed{
\boldsymbol{E}(\boldsymbol{r}) = \left( \boldsymbol{\mathscr{E}}(\boldsymbol{r}) \right)_{unif.} + \frac{1}{2\pi}  \sum_{n=1}^N (V_n - V_{n+1}) \left[\boldsymbol{f}(\beta_n,\boldsymbol{r}) + \frac{R}{\left(R^2+r^2\right)^{3/2}}\sum_{s=0}^{\infty} (-1)^s \binom{-3/2}{s} \xi^s \boldsymbol{L}_n^{(s)}(\beta_n,\boldsymbol{r}) \right]
}
\label{EVectorFieldStairCaseVExpansionEq}
\end{equation}
where $\left( \boldsymbol{\mathscr{E}}(\boldsymbol{r}) \right)_{unif.} = (\mathscr{E}_{r}(\boldsymbol{r}))_{unif.} \hat{r} + (\mathscr{E}_{\theta}(\boldsymbol{r}))_{unif.} \hat{\theta}$ and $z>0$. Finally, if $V(\phi)$ is assumed as a continuous and fully periodic function, then we may write  

\[
\boldsymbol{E}(\boldsymbol{r}) = \left( \boldsymbol{\mathscr{E}}(\boldsymbol{r}) \right)_{unif.} - \frac{1}{2\pi}  \int_{0}^{2\pi} \partial_\phi V(\beta) \left[\boldsymbol{f}(\beta,\boldsymbol{r}) + \frac{R}{\left(R^2+r^2\right)^{3/2}}\sum_{n=0}^{\infty} (-1)^s \binom{-3/2}{s} \xi^s \boldsymbol{L}_n^{(s)}(\beta,\boldsymbol{r}) \right]d\beta .
\]

Since $\boldsymbol{L}_n^{(s)}(\beta,\boldsymbol{r})$ is periodic with respect $\beta$ then
\[
\boldsymbol{E}(\boldsymbol{r}) = \left( \boldsymbol{\mathscr{E}}(\boldsymbol{r}) \right)_{unif.} + < V \boldsymbol{h} > 
\]
with 
\[
< V \boldsymbol{\kappa} > = \frac{1}{2\pi} \int_{0}^{2\pi} V(\beta) \partial_\beta \boldsymbol{\kappa}(\beta,\boldsymbol{r}) d\beta
\]
and
\[
\boldsymbol{\kappa}(\beta,\boldsymbol{r}) =  \boldsymbol{f}(\beta,\boldsymbol{r}) + \frac{R}{\left(R^2+r^2\right)^{3/2}}\sum_{n=0}^{\infty} (-1)^s \binom{-3/2}{s} \xi^s \partial_\phi\boldsymbol{L}_n^{(s)}(\beta,\boldsymbol{r}). 
\]

In practice formula Eq.~(\ref{EVectorFieldStairCaseVExpansionEq}) requires to compute the complete elliptic integral implicit in $\left( \boldsymbol{\mathscr{E}}(\boldsymbol{r}) \right)_{unif.}$. To this aim we may use their series representations \cite{radon1950sviluppi}
\begin{align*}
\b{E}(\chi) = \frac{\pi}{2} + \frac{\pi}{2}\sum_{m=1}^\infty \left[\frac{(2m-1)!!}{(2m)!!}\right]^2\frac{1}{1-2m} \chi^{2m}, &&\mbox{and} &&K(\chi) = \frac{\pi}{2} + \frac{\pi}{2}\sum_{m=1}^\infty  \frac{e_m}{1-2m} \chi^{2m}
\end{align*}
with $n!!$ the double factorial, or use a programming package where those functions are implemented. In general, the evaluation of Eq.~(\ref{EVectorFieldStairCaseVExpansionEq}) does not offer a challenge technical problem since this analytic expression can be coded into a high-level language program. In particular, we have written a short notebook in Wolfram Mathematica 9.0 \cite{wolfram2012version} in order to explore Eq.~(\ref{EVectorFieldStairCaseVExpansionEq}) for an arbitrary potential $V(\phi)$. We have chosen $\mathcal{V}(\phi)$ as follows 
\begin{equation}
\mathcal{V}(\phi) = U_o\left\{\frac{1}{5} + \exp\left[-(\phi-\pi)^2\right]\right\}
\label{gaussianContinousVEq}    
\end{equation}

with $U_o=1$, however other selection of $\mathcal{V}(\phi)$ is also valid if it fulfills with $\mathcal{V}(0)=\mathcal{V}(2\pi)$. On Fig.~\ref{EComponentsSurfFig} we show some surfaces corresponding to the electric field components for $\theta = 2\pi/5$, relatively near to the $xy$-plane where $\theta \rightarrow \pi/2$. Plots in  Fig.~\ref{EComponentsSurfFig} were obtained with a staircase like potential $V(\phi)$ of $N=7$ sectors shown in Fig.~\ref{discretePotentialLimitFig}-(left). Such potential introduces introduces $N-1$ discontinuities which can affect drastically the electric field near the $xy$-plane, as we can observe on Fig.~\ref{EComponentsSurfFig}. In general, these sudden changes on the electric field components with respect the $\phi$-coordinate and near the $xy$-plane must disappear in the limit $N \rightarrow \infty$ if $\mathcal{V}(\phi)$ is selected as a continuous smooth function. We may start to observe this behaviour by choosing a value of $N$ relatively large (see  Fig.~\ref{EComponentsSurfIIFig}) where $N=33$. Finally, some plots of the vector field are shown in Fig.~\ref{vectorFieldEPhiConstFig}.    

\section{Numerical comparison}

In this document we described an analytic technique to compute the electric field due to a planar region hold at some potential $V(\phi)$. 

\begin{figure}[H]
\centering
\includegraphics[width=0.325\textwidth]{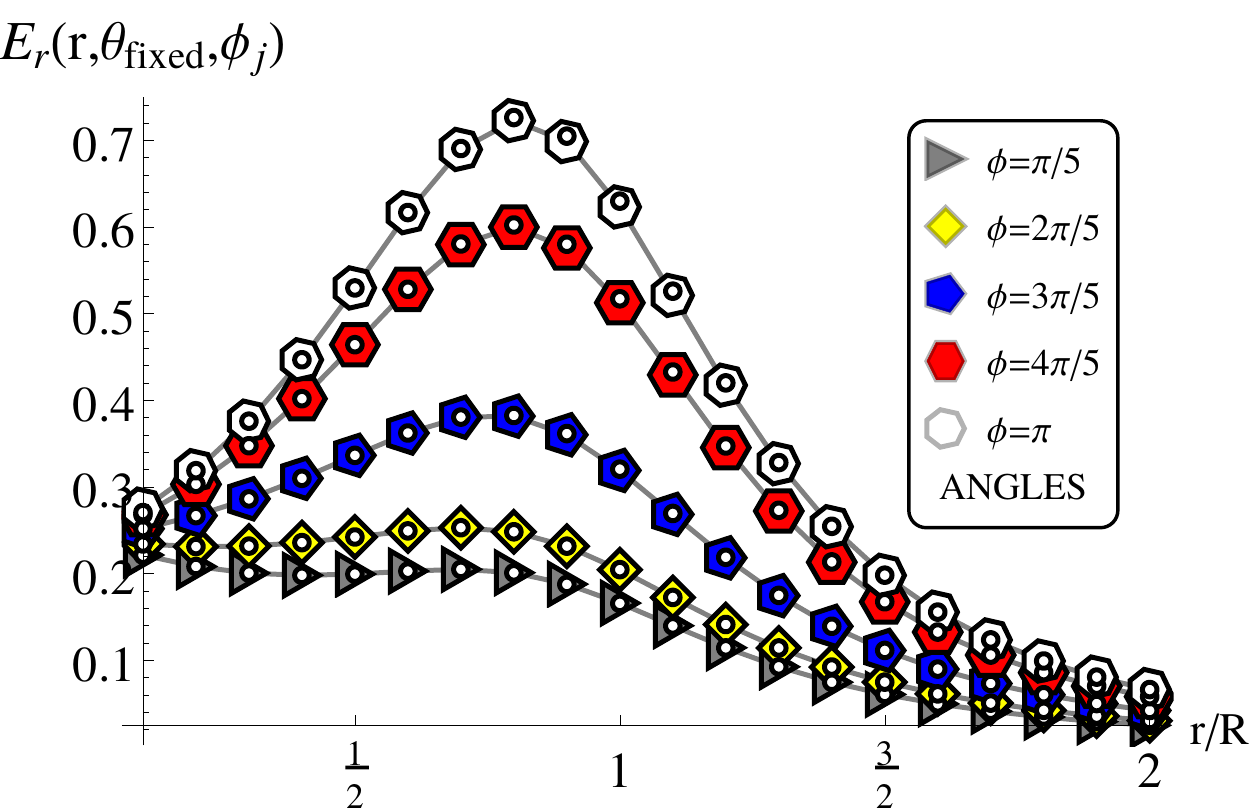}%ErComparison.pdf
\includegraphics[width=0.325\textwidth]{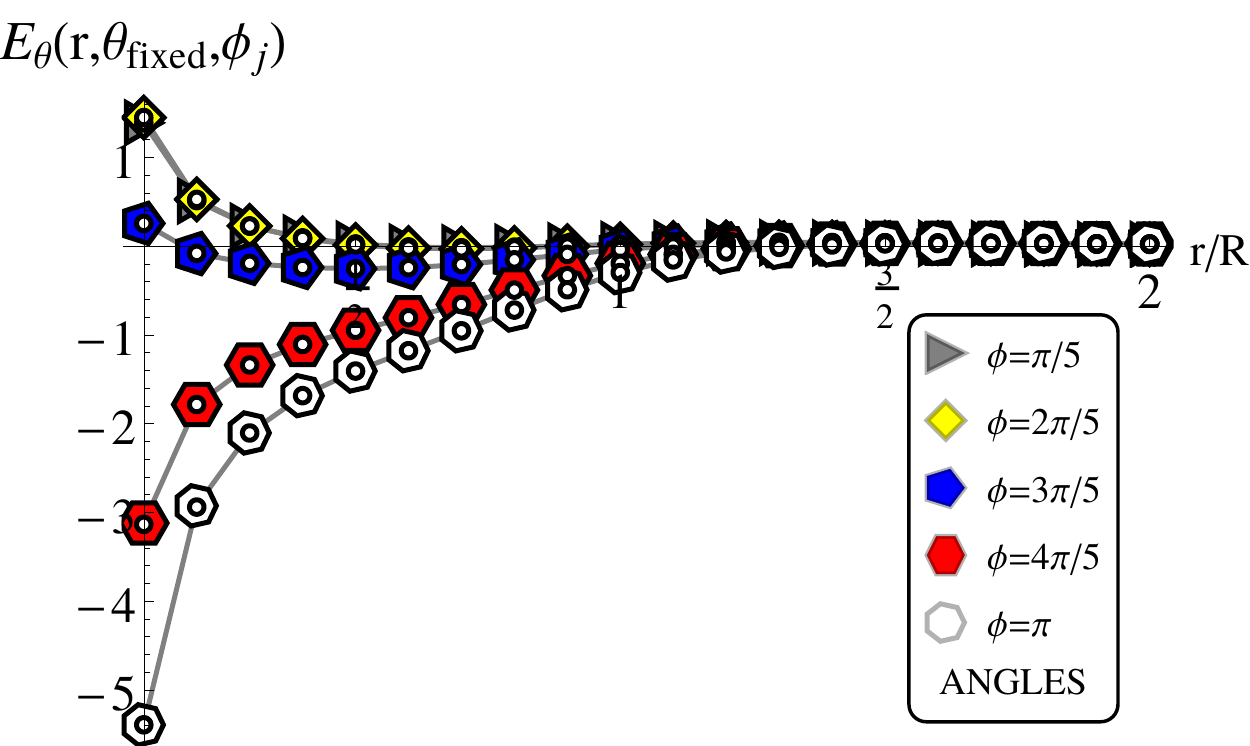}%EThetaComparison.pdf
\includegraphics[width=0.325\textwidth]{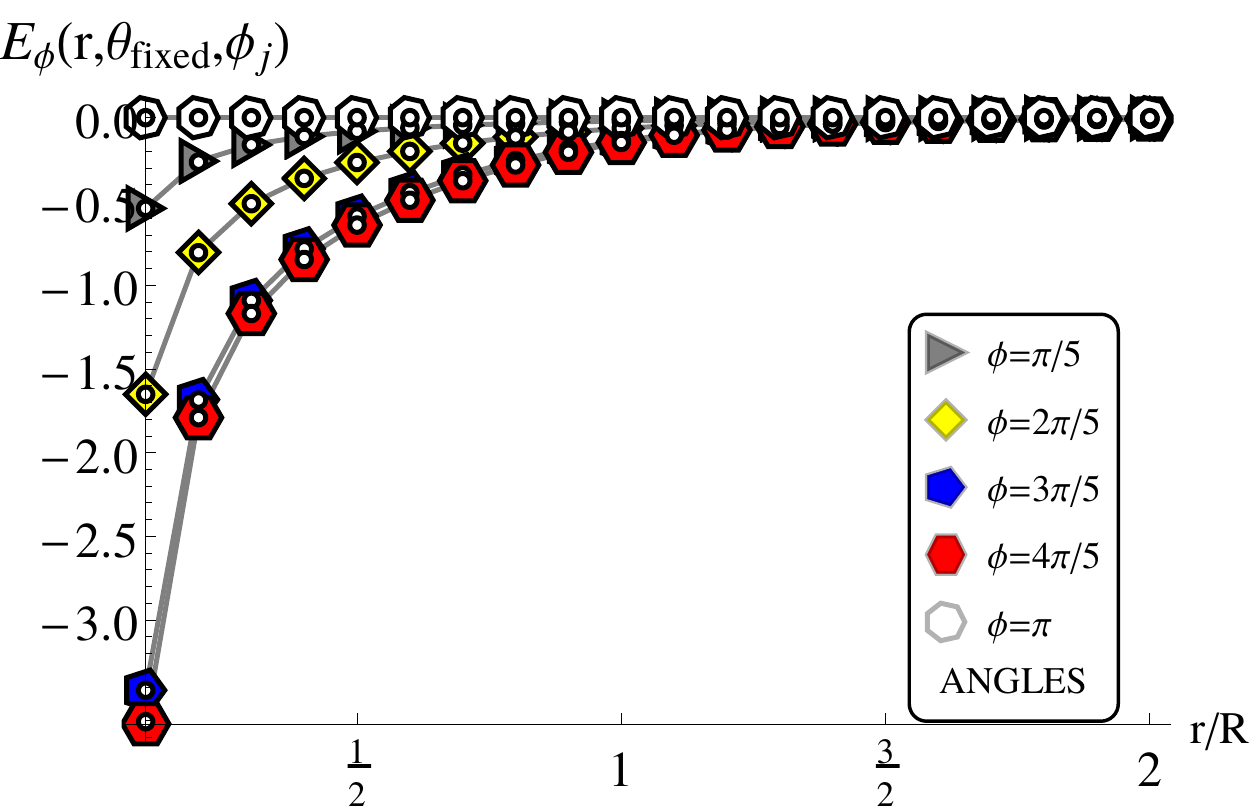}%EPhiComparison.pdf
    \caption{Numeric and series expansions of $\boldsymbol{E}(r,\theta,\phi)$ at $\theta = \pi/3$. (From left) Components of the electric field. Polygonal symbols are the electric field computed with Eq.~(\ref{EVectorFieldStairCaseVExpansionEq}) and open dots correspond to numerical integration Eq.~(\ref{electricPotentialDoubleIntegralEq}) plus a central differentiation corresponding to the gradient components. We have used $M=20$ terms in the series. Gray solid lines connecting symbols in this plots are only  included to guide the eye.}
\label{EComparisonPhipi3Fig}
\end{figure}

However, there are also numerical alternatives to solve this problem. One of them is to evaluate numerically the double integral in Eq.~(\ref{electricPotentialDoubleIntegralEq}) and apply a numerical central differentiation on this integral to compute an approximated gradient \cite{brezillon1981numerical,burden2001numerical}, this is
\[
E_r(\boldsymbol{r}) \approx - \frac{1}{h} \left[\Phi\left(r+\frac{h}{2},\theta,\phi\right)-\Phi\left(r-\frac{h}{2},\theta,\phi\right)\right],
\]
\[
E_\theta(\boldsymbol{r}) \approx - \frac{1}{r h} \left[\Phi\left(r,\theta+\frac{h}{2},\phi\right)-\Phi\left(r,\theta-\frac{h}{2},\phi\right)\right]
\]
and
\[
E_\phi(\boldsymbol{r}) \approx - \frac{1}{r \sin\theta h} \left[\Phi\left(r,\theta,\phi+\frac{h}{2}\right)-\Phi\left(r,\theta,\phi-\frac{h}{2}\right)\right].
\]

To this aim, we used the function NIntegrate of Wolfram Mathematica 9.0 \cite{wolfram2012version} to evaluate $\Phi\left(r,\theta,\phi\right)$ and their shifted values with $h$ small. Even when this strategy is easy to implement, there are some sources of error that must be taken into account. The first one emerges from the numerical integration since the integrand in Eq.~(\ref{electricPotentialDoubleIntegralEq}) may become highly oscillatory depending on the potential function $V(\phi)$ and the point of evaluation. Another source of error comes from the numerical differentiation. Typically, the error of central derivatives is proportional to $h^2$, however in numerical computations $h$ cannot be set arbitrarily small  because the numerator of the finite derivative can be cancelled \cite{squire1998using}. This error depends on the machine precision.

\begin{figure}[h]
\centering%analyticErFunctionOfNPhi2pi5.pdf
\includegraphics[width=0.4\textwidth]{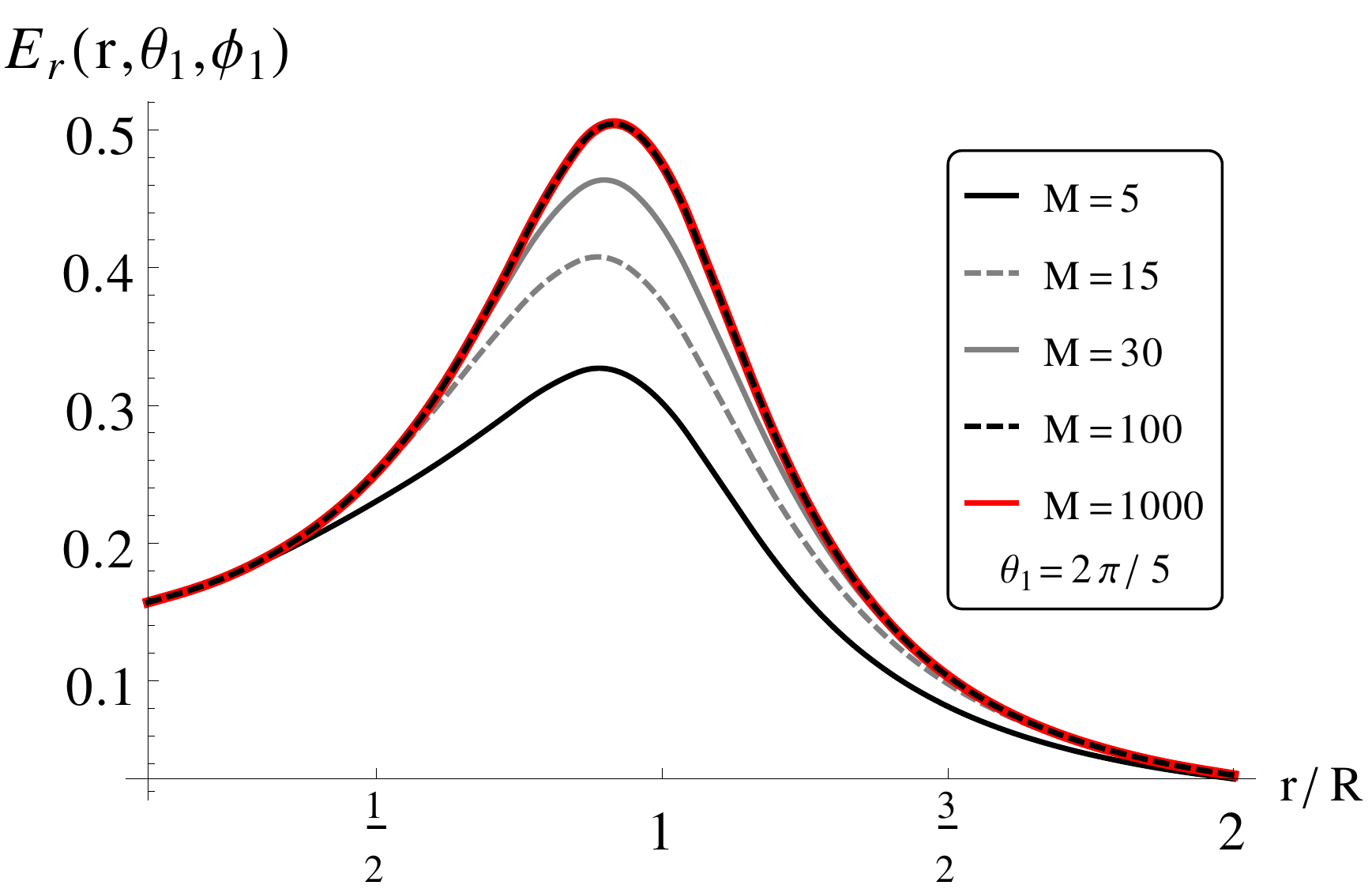}\hspace{0.5cm}%analyticErFunctionOfNPhipi3.pdf
\includegraphics[width=0.4\textwidth]{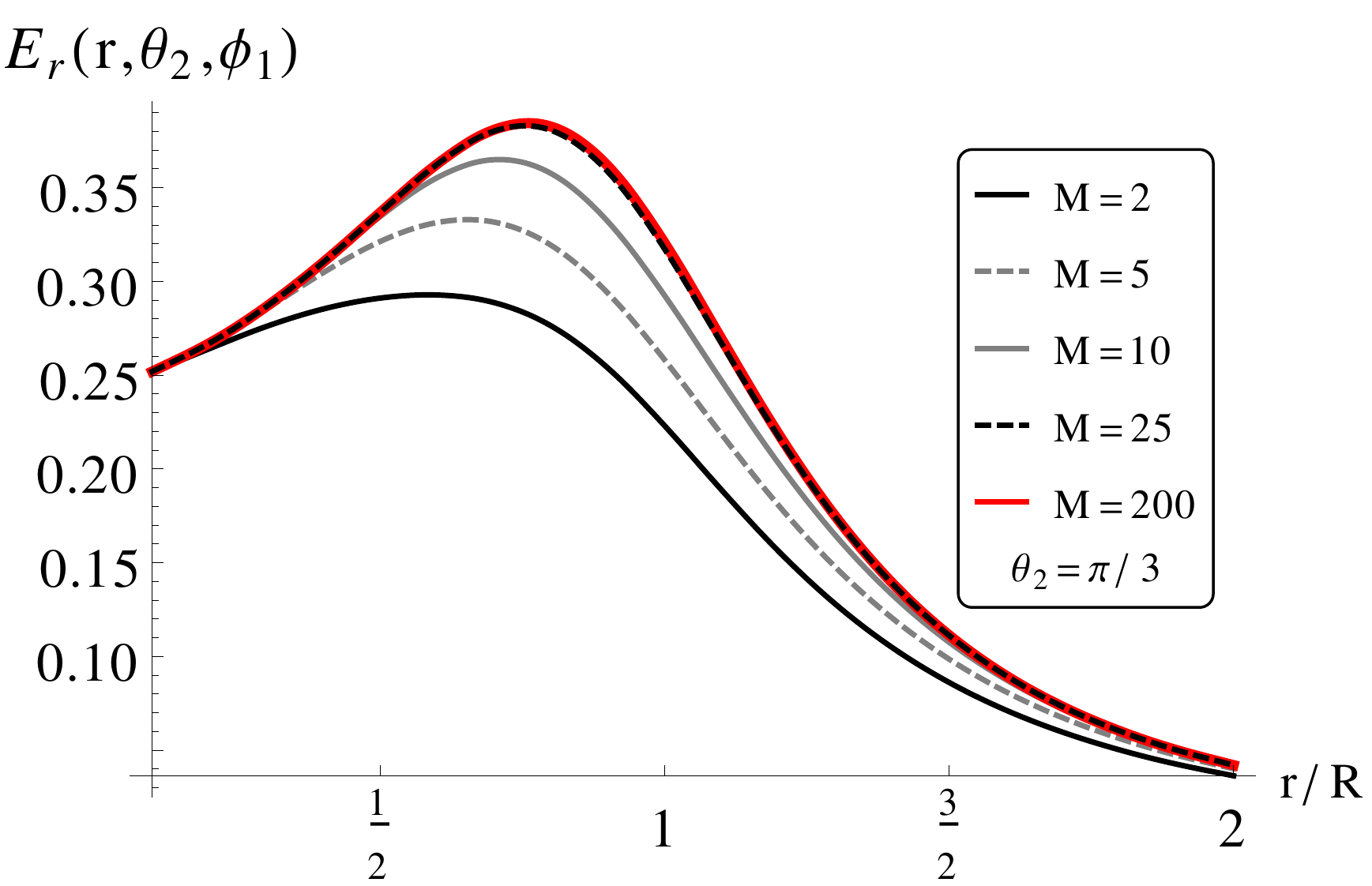}
    \caption{Series solution. Radial electric field at (left) $\theta_1=2\pi/5$ and (right) $\theta_2=\pi/3$. The value of $\phi$ was fixed at $\phi_1=\pi/5$ in both plots. }
\label{EComparisonNBehaviourFig}
\end{figure}

\begin{figure}[H]
\centering %ERROR.pdf
\includegraphics[width=0.45\textwidth]{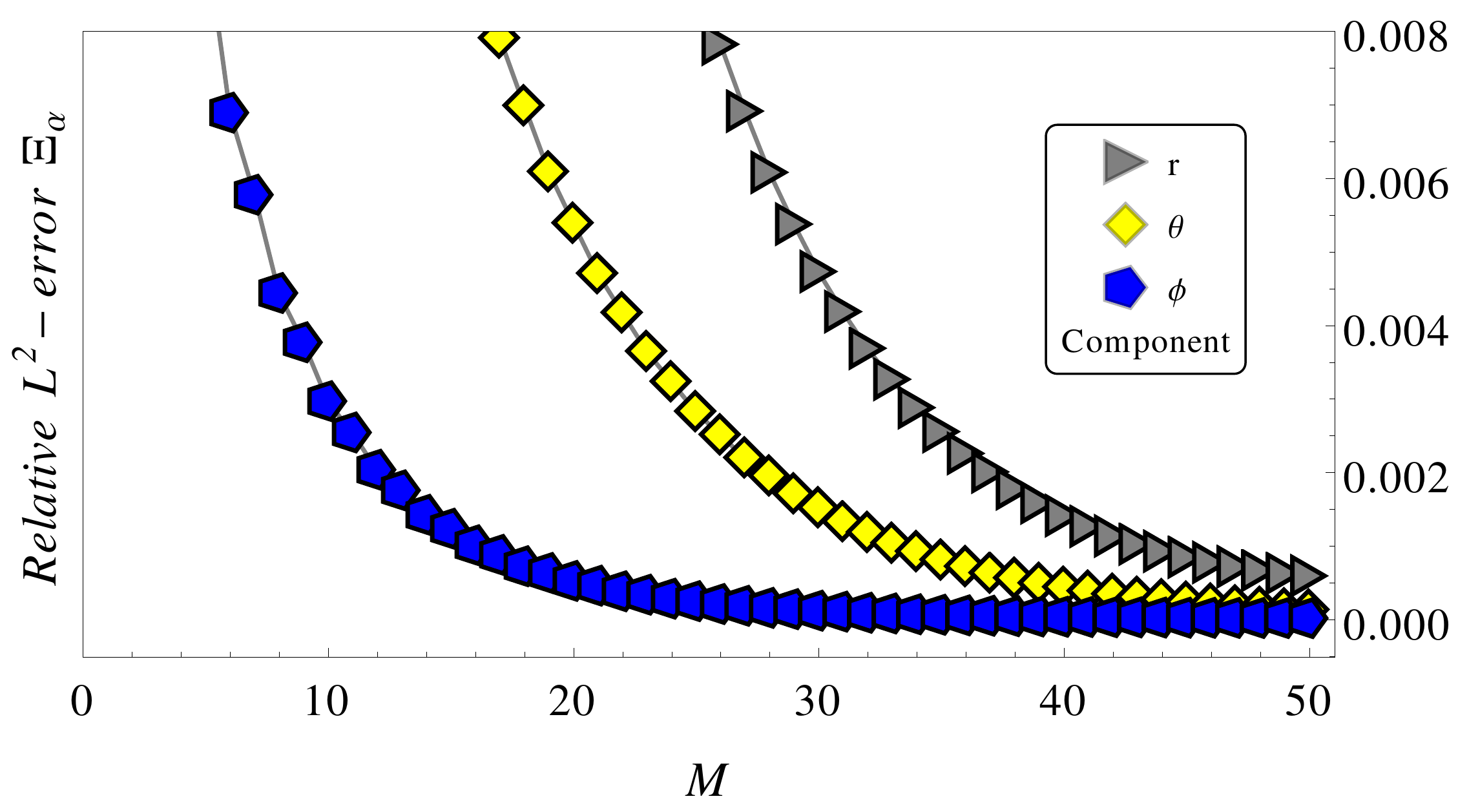}
    \caption{$L^2$ Relative error norm. }
\label{EErrorFig}
\end{figure}

On Fig.~\ref{EComparisonPhipi3Fig}, we show a computation  of the electric field by using numeric integration and a partial differentiation. The potential function $V(\phi)$ was chosen as a staircase-like function with $N=33$ and $\mathcal{V}$ Gaussian as it is shown in Fig.~\ref{discretePotentialLimitFig}-(center). We also show on Fig.~\ref{EComparisonPhipi3Fig} the analytic values of the electric field according to Eq.~(\ref{EVectorFieldStairCaseVExpansionEq}) at $\theta=\pi/3$ for few values of $\phi$ in $[0,\pi]$. Numerical and analytic computations of the electric field components are in agreement. In general, the evaluation of the analytic expression requires a truncation of the infinite sum at the right of Eq.~(\ref{EVectorFieldStairCaseVExpansionEq}). Such infinite sum comes from the Taylor series of Eq.~(\ref{binomialTheoremSeriesExpansionEq}) which converges for $|\chi|<1$ for any complex number $\alpha$. However, if the series is truncated at the first $M$ terms, then we have to increase $M$ as $|\chi|$ approaches to one in order to obtain descent results with a small error. This error dependence of the truncated series with $M$ is inherited on the computation of the computation of the electric field Eq.~(\ref{EVectorFieldStairCaseVExpansionEq}).  Since $\chi(\boldsymbol{r},\boldsymbol{r}') = \frac{2 R r \sin\theta}{R^2+r^2} \cos(\phi-\phi')$ then we require more series terms when we try to evaluate the electric field on the $xy$-plane (this is at $\theta=\pi/2$) and near $r=R$ since $\chi$ approaches to one if $\phi'\rightarrow\phi$. This dependence can be observed on Fig.~(\ref{EComparisonNBehaviourFig}) where we show the behaviour of the analytic solution as function of $M$ at $\theta = 2\pi/5$ and $\theta = \pi/3$. We require at least $M=100$ terms for $\theta = 2\pi/5$. On the other hand, $M=25$ terms are enough to compute $E_r$ at $\theta = \pi/3$. For this reason, we were able to use $M=20$ terms on the series computations shown in Fig.~\ref{EComparisonPhipi3Fig} without having a large difference between the truncated series and the values via numeric integration and numeric differentiation even when $M$ is small. In order to find the error estimates on the electric field components, we calculated the following $L^2$ relative error norm
\[
\Xi_{\alpha} = \sqrt{\frac{1}{\sum_{(r_i,\theta_j,\phi_k)\in\mathcal{D}}  E_{\alpha}^2( \boldsymbol{r}_{ijk} )}\sum_{(r_i,\theta_j,\phi_k)\in\mathcal{D}} \Delta E_{\alpha}^2( \boldsymbol{r}_{ijk} ) } 
\]
for the  $\alpha$-th component of the electric field in spherical coordinates. Here $\Delta E_{\alpha} = E_{\alpha} - E_{\alpha}^{(num.)}$ where $E_{\alpha}$ and $E_{\alpha}^{(num.)}$ are the components of $\boldsymbol{E}$ computed with Eq.~(\ref{EVectorFieldStairCaseVExpansionEq}) and numerically respectively. The set $\mathcal{D}$ was defined as rectangular cuboid of volume $2R \times \theta_{max} \times 2\pi$ where $r\in [0,2R]$, $\theta\in [0,\theta_{max}]$ and  $\phi \in [0, 2\pi)$. The value of $\theta_{max}$ was set as $0.8\pi/2$, the total points in the rectangular lattice of indices $\{(i,j,k)\}$ was 8000. A plot of the $L^2$ relative error norm as a function of number of terms in Eq.~(\ref{EVectorFieldStairCaseVExpansionEq}) is shown in Fig.~\ref{EErrorFig}.  

\subsection{FEM}

Consider the spatial domain $\Omega$, with boundary $\partial\Omega$ and $\boldsymbol{n}$ the unit normal to the boundary.
The weak form of the Laplace equation for the potential field inside $\Omega$ is readily obtained by integrating it by parts and using a test function $v\in V$, with $V$ being a suitable function space that satisfies the Dirichlet boundary conditions (i.e. the potential distribution over the planar region):
\[
\int_{\Omega} \boldsymbol{\nabla}u\cdot\boldsymbol{\nabla}v \, {\rm d} x = 
 \int_{\partial\Omega} (\boldsymbol{\nabla}u\cdot \boldsymbol{n}) v \, {\rm d} s \quad \text{for all } v\in V.
\]

\begin{figure}[H]
\centering%errorFEM.pdf
\includegraphics[width=0.45\textwidth]{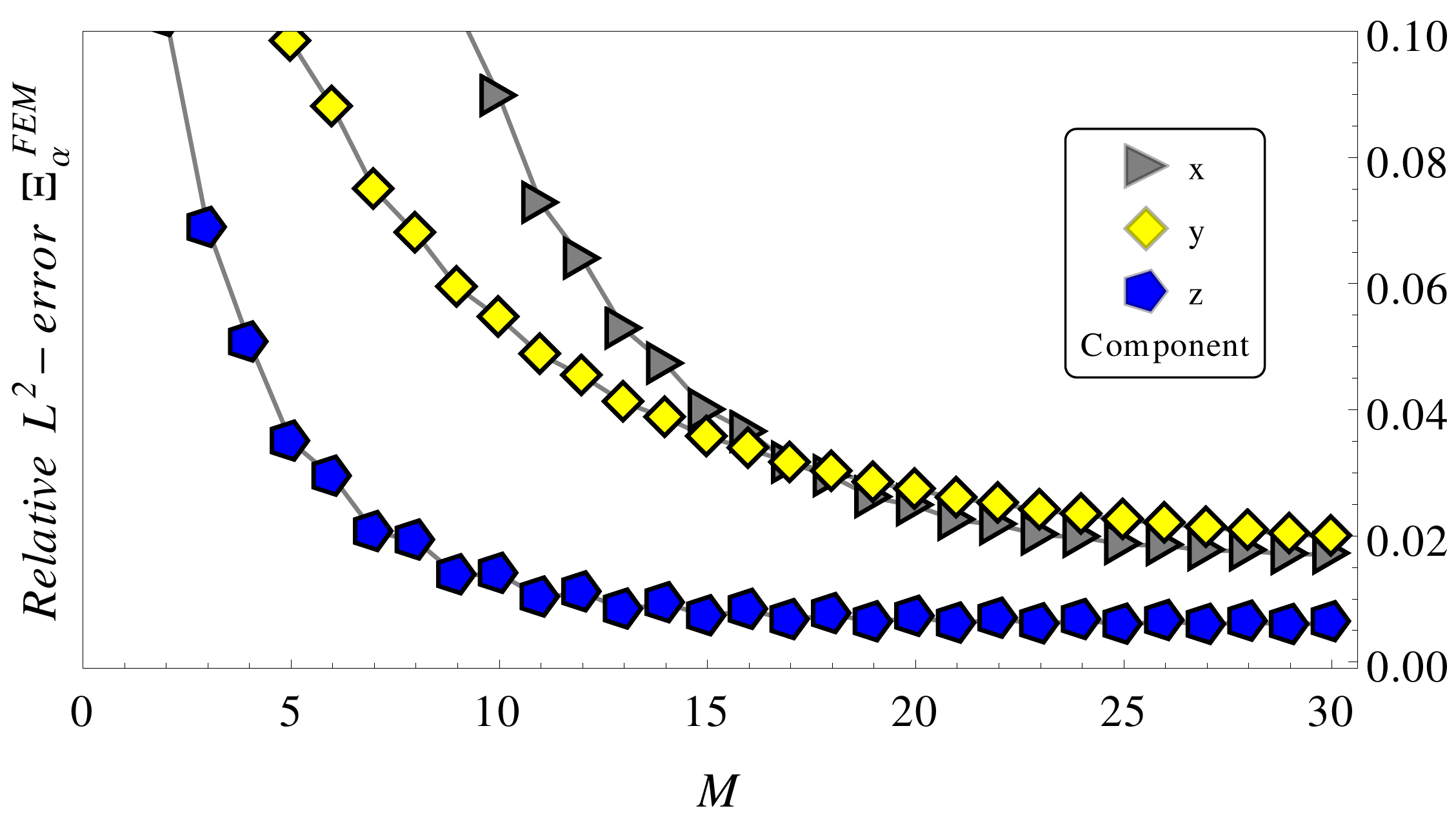}
    \caption{$L^2$ Relative error norm. Numerical values were computed via FEM. }
\label{EErrorFEMFig}
\end{figure}

In order to formulate the Galerkin approximation of the Laplace's problem, let us first denote by $\mathcal{T}_h =\left\{\Omega^e\right\}$ the finite element partition of the domain $\Omega$, with index $e$ ranging from 1 to the number of elements $n_{el}$ in the finite mesh.
The diameter of the element partition is denoted by $h$.
We define the finite test function spaces as made of continuous piecewise polynomial functions in space.
The Galerkin approximation of the weak Laplace's problem considers replacing $V$ by the finite subspaces $V_h\subset V$, where the subscript $h$ refers the discrete finite element space.
The problem to be solved now is to find $u_h,\in V_h$ such that 
\begin{equation}
\int_{\Omega} \boldsymbol{\nabla}u_h\cdot\boldsymbol{\nabla}v_h \, {\rm d} x = 
 \int_{\partial\Omega} (\boldsymbol{\nabla}u_h\cdot \boldsymbol{n}) v_h \, {\rm d} s \quad \text{for all } v_h\in V_h. \label{FEM}    
\end{equation}{}

\begin{figure}[ht]  
  \begin{minipage}[b]{0.45\linewidth}
    
   \centering 
   \includegraphics[width=1\textwidth]{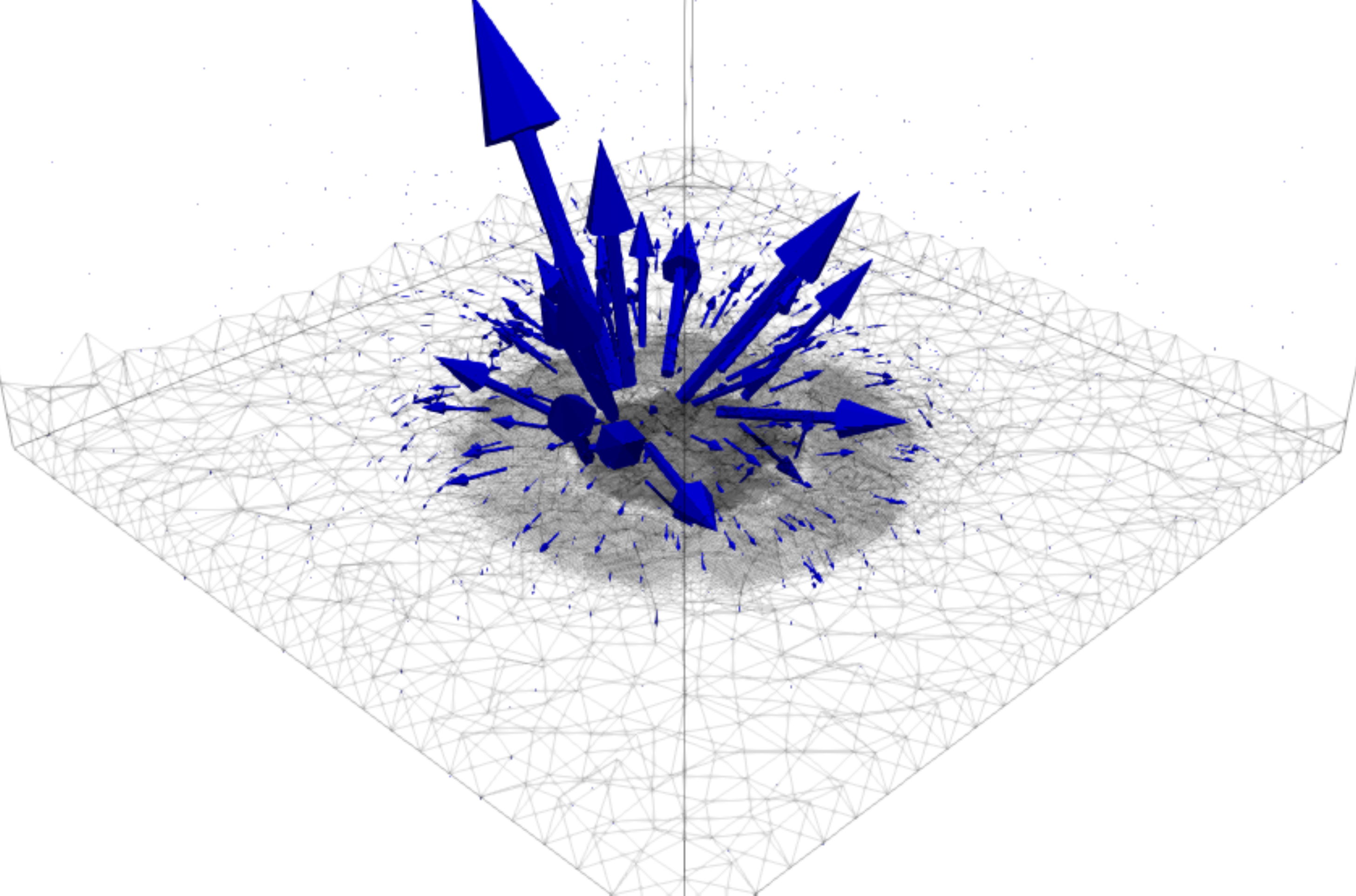}
   \caption*{(a)} %cropFig.pdf
    
  \end{minipage} 
  \begin{minipage}[b]{0.5\linewidth}
    
    \includegraphics[width=1.01\textwidth,trim = 3cm 1.5cm 3cm 9cm, clip=true]{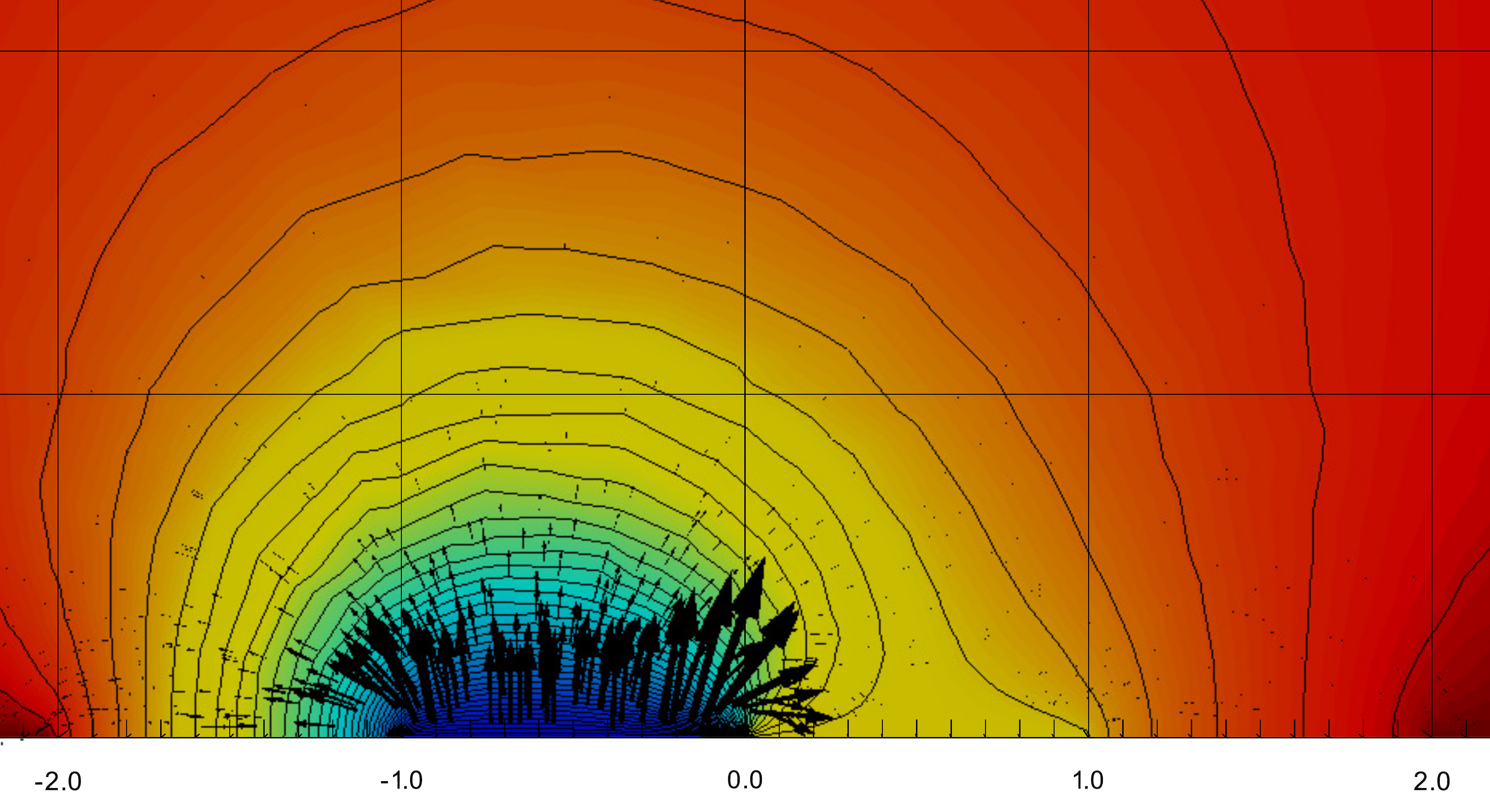}
    \caption*{(b) $\phi=0$} %FEMPhi0.eps
  
  \end{minipage} 
  \begin{minipage}[b]{0.5\linewidth}
    
    \includegraphics[width=1\textwidth,trim = 3cm 1.5cm 3cm 9cm, clip=true]{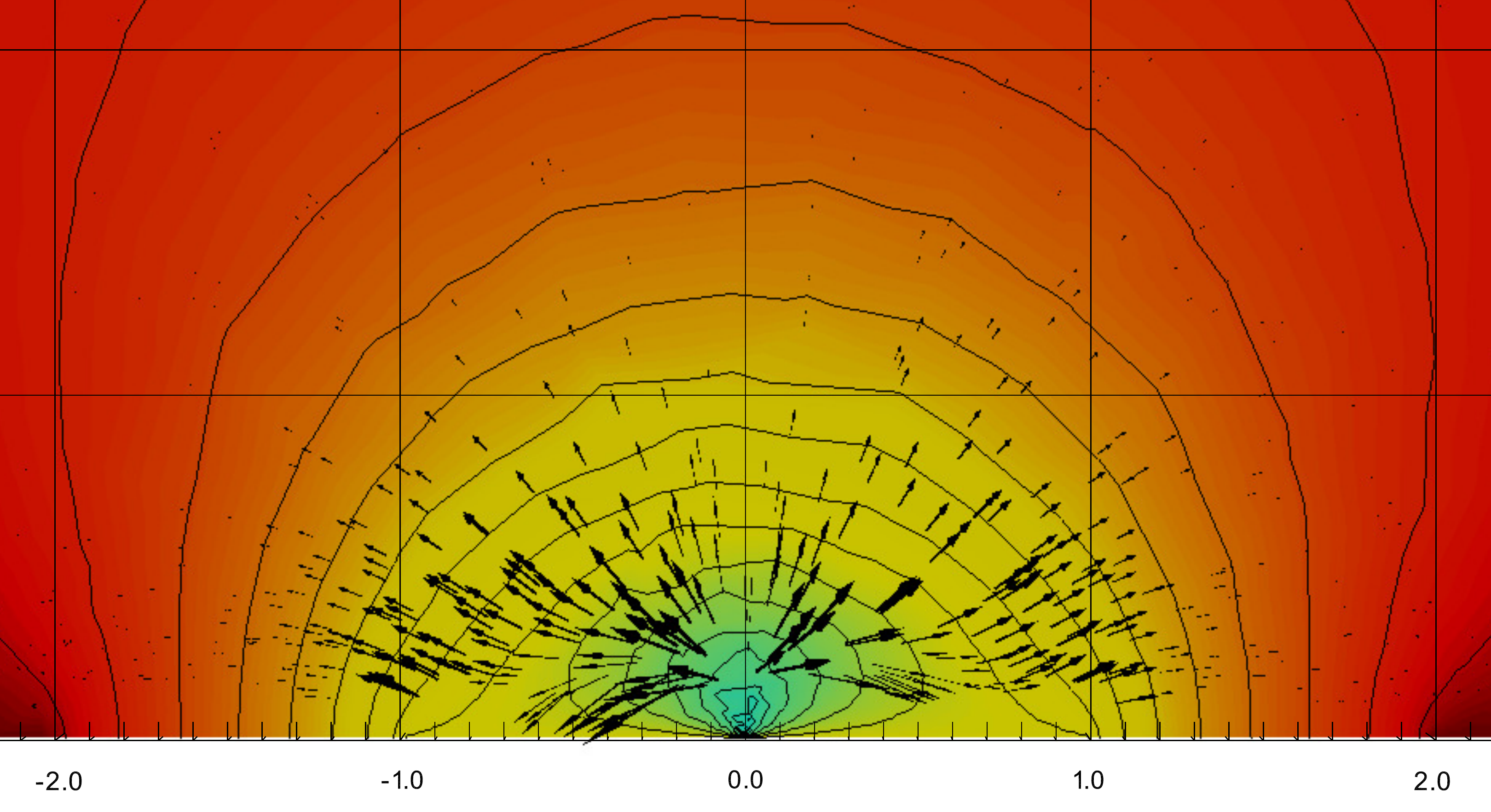}
    \caption*{(c) $\phi=\pi/2$}  %FEMPhi180.eps
  
  \end{minipage}
  \hfill
  \begin{minipage}[b]{0.5\linewidth}
    \includegraphics[width=1\textwidth,trim = 3cm 1.5cm 3cm 9cm, clip=true]{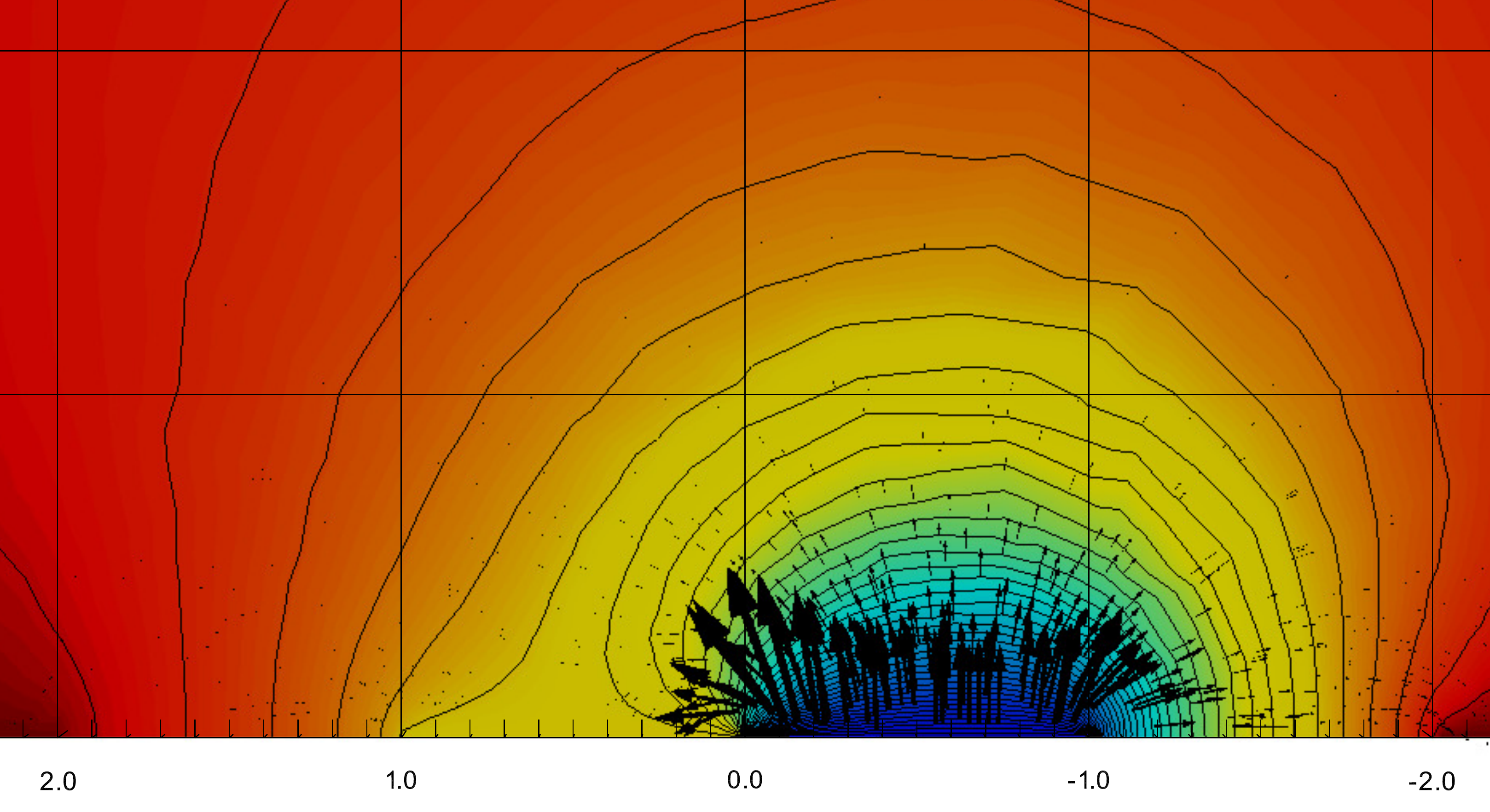}
    \caption*{(d) $\phi=3\pi/4$} %FEMPhi270.eps
    
  \end{minipage}
  
  \caption{ Electric field. (a) Numerical vector field computed with Eq.~(\ref{FEM}). Computations were performed for a Gaussian potential $\mathcal{V}$ (see Eq.~(\ref{gaussianContinousVEq})) using a $n_{el}=439014$ linear tetrahedron unstructured mesh with local refinement near the potential region. (b), (c) and (d) are projections of the $\boldsymbol{E}(\boldsymbol{r})$ on the plane $\phi=0, \pi/2$ and $3\pi/4$ respectively.  }
  
  \label{FEMFig}
  
\end{figure}

We obtained the results displayed in Fig.~\ref{FEMFig} for linear tetrahedron type of finite elements.
Specifically, we used an unstructured mesh composed by $n_{el}=439014$ to discretize the prismatic domain $\Omega:=[-4R,4R] \times [-4R,R] \times [0,8R]$, with local refinement near the planar circular region of radius $R$ that is centered on one of the boundaries. On Fig.~\ref{EErrorFEMFig} we show the $L^2$ Relative error norm between the Eq.~(\ref{EVectorFieldStairCaseVExpansionEq}) and the numerical results of FEM. Solutions approach each other as $M$ is increased. Even when the current problem can be solved numerically, we stress the fact that truncated analytic series from Eq.~(\ref{EVectorFieldStairCaseVExpansionEq}) can be computationally less expensive than traditional numerical approaches if $\theta$ is not near to $\pi/2$. This is just the behaviour of analytic solution shown in  Fig.~\ref{EComparisonNBehaviourFig} where the truncated series tends faster to the exact result as $\theta$ is decreased from $\pi/2$.

\section*{Conclusion}
In this work we presented an approach to compute the electric field due to planar regions kept at a fixed but non-uniform potential $V(\phi)$. We used some connections of this problem with magnetostatics to simplify the electrostatic problem. As we shown in Eq.~(\ref{electricFieldBiotSavartWithCorrectionEq}), the electric field can be found by evaluation of two one-dimensional integrals depending on $\phi$, where one of them is analogous to the well-known Biot-Savart law of magnetostatics. Using this approach, it is possible to find an exact series solution (see Eq.~(\ref{EVectorFieldStairCaseVExpansionEq})) of the problem when a circular region is considered.

\section*{Acknowledgments}
This work was supported by Vicerrector\'ia de investigaci\'on, Universidad ECCI. Robert Salazar also thanks Fundaci\'on Colfuturo and Departamento de Ciencias B\'asicas, Universidad ECCI.
%----------------------------------------------------------------------------------------
%	REFERENCE LIST
%----------------------------------------------------------------------------------------
\bibliographystyle{ieeetr} %alpha, apalike, ieeetr
\bibliography{bibliography.bib}

%\begin{thebibliography}{99} % Bibliography - this is %intentionally simple in this template

%\bibitem{andreotti1997studying} B. Andreotti, \emph{Studying Burgers' models to investigate the physical meaning of the alignments statistically observed in turbulence}, Phys. Fluids \textbf{9} : 3, March (1997)

%\bibitem{cohl1999compact} Cohl, Howard S., and Joel E. Tohline, \emph{A compact cylindrical Green's function expansion for the solution of potential problems}, The astrophysical journal \textbf{527} : 86 - 101 (1999) %DOI: https://doi.org/10.1086/308062

%\bibitem{abramowitz1965handbook} Abramowitz, Milton, and Irene A. Stegun. \emph{Handbook of Mathematical Functions With Formulas, Graphs, and Mathematical Tables.} (1964).

%\end{thebibliography}

%croos section
%GOOD: http://www.eumetrain.org/satmanu/CMs/TrCyAt/print.htm
%https://physics.stackexchange.com/questions/275799/why-is-the-eye-of-a-cyclone-a-forced-vortex
%http://www.chanthaburi.buu.ac.th/~wirote/met/tropical/textbook_2nd_edition/navmenu.php_tab_9_page_7.1.0.htm
%http://www.atmos.umd.edu/~dalin/andrew/part2.html
%https://nptel.ac.in/courses/119102007/2
%http://www.911omissionreport.com/steering_hurricanes.html
%https://www.youtube.com/watch?v=_brY_9ME8iE
\end{document}